\documentclass{article}

\usepackage[final]{neurips_2020}

\usepackage{todonotes}
\usepackage[utf8]{inputenc} 
\usepackage[T1]{fontenc}    
\usepackage{hyperref}       
\usepackage{url}            
\usepackage{booktabs}       
\usepackage{amsfonts}       
\usepackage{nicefrac}       
\usepackage{microtype}      
\usepackage{graphicx}
\usepackage{float}
\usepackage{multirow}
\usepackage{resizegather}
\usepackage{caption}
\usepackage{subcaption}

\usepackage{dcolumn,booktabs}

\usepackage{framed}
\usepackage[ruled, vlined]{algorithm2e}

\usepackage{amsmath,amsfonts,bm}

\def\eqref#1{equation~\ref{#1}}

\def\1{\bm{1}}

\def\rx{{\textnormal{x}}}
\def\ry{{\textnormal{y}}}

\def\rva{{\mathbf{a}}}
\def\rvb{{\mathbf{b}}}

\def\rvx{{\mathbf{x}}}

\def\rvz{{\mathbf{z}}}

\def\va{{\bm{a}}}

\def\vx{{\bm{x}}}

\def\vz{{\bm{z}}}

\DeclareMathAlphabet{\mathsfit}{\encodingdefault}{\sfdefault}{m}{sl}
\SetMathAlphabet{\mathsfit}{bold}{\encodingdefault}{\sfdefault}{bx}{n}

\newcommand{\KL}{D_{\mathrm{KL}}}

\DeclareMathOperator*{\argmax}{arg\,max}

\DeclareMathOperator*{\argtop}{arg\,top}

\newcommand{\X}{\mathcal{X}}

\newcommand{\Oh}{\mathcal{O}}
\renewcommand{\KL}[2]{\mathrm{KL}\left[\,#1\,\,||\,\,#2\,\right]}
\newcommand{\KLqp}{\KL{q(\rvz\mid\vx)}{p(\rvz)}}
\newcommand{\Exp}[2]{\mathbb{E}_{#2}\left[ #1 \right]}

\newcommand{\Norm}[1]{\mathcal{N}\left( #1 \right)}

\renewcommand{\mid}{\,|\,}

\newcommand{\widesim}[2][1.5]{
  \mathrel{\overset{#2}{\scalebox{#1}[1]{$\sim$}}}
}

\title{Compressing Images by Encoding Their Latent Representations with Relative Entropy Coding}

\author{%
  Gergely Flamich\thanks{Equal contribution.} \\
  Department of Engineering\\
  University of Cambridge\\
  \texttt{gf332@cam.ac.uk} \\
  \And
  Marton Havasi\footnotemark[1]\\
  Department of Engineering\\
  University of Cambridge\\
  \texttt{mh740@cam.ac.uk} \\
  \And
  Jos\'e Miguel Hern\'andez-Lobato \\
  Department of Engineering\\
  University of Cambridge, \\
  Microsoft Research, \\
  Alan Turing Institute\\
  \texttt{jmh233@cam.ac.uk} \\
}

\begin{document}

\maketitle

\begin{abstract}
Variational Autoencoders (VAEs) have seen widespread use in learned image compression. They are used to learn expressive latent representations on which downstream compression methods can operate with high efficiency.
Recently proposed `bits-back' methods can indirectly encode the latent representation of images with codelength close to the relative entropy between the latent posterior and the prior. 
However, due to the underlying algorithm, these methods can only be used for lossless compression, and they only achieve their nominal efficiency when compressing multiple images simultaneously; they are inefficient for compressing single images.
As an alternative, we propose a novel method, Relative Entropy Coding (REC), that can directly encode the latent representation with codelength close to the relative entropy for single images, supported by our empirical results obtained on the Cifar10, ImageNet32 and Kodak datasets. Moreover, unlike previous bits-back methods, REC is immediately applicable to lossy compression, where it is competitive with the state-of-the-art on the Kodak dataset.
\end{abstract}

\section{Introduction}
\par 
The recent development of powerful generative models, such as Variational
Autoencoders (VAEs) has caused a great deal of interest in their application to image compression, notably 
\cite{balle2016end, balle2018variational, townsend2020hilloc, minnen2020channel}.
The benefit of using these models as opposed to hand-crafted methods is that
they can adapt to the statistics of their inputs more effectively, and hence allow
significant gains in compression rate. A second advantage is their easier adaptability to
new media formats, such as light-field cameras, 360$^\circ$ images, Virtual
Reality (VR), video streaming, etc. for which classical methods are not
currently applicable, or are not performant.

\par
VAEs consist of two neural networks, the encoder and the decoder. The former maps images to their latent representations and the latter maps them back. Compression methods can operate very efficiently in latent space, thus realizing a non-linear transform coding method \citep{goyal2001theoretical, balle2016endtrans}. The sender can use the encoder to obtain the latent posterior of an image and then use the compression algorithm to transmit a sample latent representation from the posterior. Then, the receiver can use the decoder to reconstruct the image from the latent representation they received. Note that this reconstruction contains small errors. In lossless compression, the sender must correct for this and has to transmit the residuals along with the latent code. In lossy compression, we omit the transmission of the residuals, as the model is optimized such that the reconstruction retains high perceptual quality.

\par
Bits-back methods for lossless compression have been at the center of attention recently. They realize the optimal compression rate postulated by the bits-back argument \citep{hinton1993keeping}: For a given image, the optimal compression rate using a latent variable model (such as a VAE) is given by the relative entropy between the latent posterior and the prior $\KLqp$ \textit{plus} the expected residual error $\Exp{-\log P(\vx \mid \rvz)}{}$, where $\vx$ is the input image, $\rvz$ denotes the stochastic latent representation, and $p(\rvz)$ is the prior over the latent space.
This quantity is also known as the negative Evidence Lower BOund (ELBO).

\par
Current bits-back compression methods use variants of the Bits-Back with Asymmetric Numeral Systems (BB-ANS) algorithm \citep{townsend2019practical, townsend2020hilloc, ho2019compression, kingma2019bit}. BB-ANS can achieve the bits-back compression rate asymptotically by allowing the codes in a sequence of images to overlap without losing information (hence getting the bits back). The issue is that the first image requires a string of auxiliary bits to start the sequence, which means that it is inefficient when used to compress a single image. Including the auxiliary bits, the compressed size of a single image is often 2-3 times the original size.\footnote{Based on the number of auxiliary bits recommended by \citet{townsend2020hilloc}.}

\par
We introduce Relative Entropy Coding (REC), a lossless compression paradigm that subsumes bits-back methods. A REC method can encode a sample from the latent posterior $\vz \sim q(\rvz\mid\vx)$ with codelength close to the relative entropy $\KLqp$, given a shared source of randomness. Then the residuals are encoded using an entropy coding method such as arithmetic coding \citep{witten1987arithmetic} with codelength $-\log P(\vx\mid\vz)$. This yields a combined codelength  close to the negative ELBO in expectation without requiring any auxiliary bits.

\par
We propose a REC method, called index coding (iREC), based on importance sampling. To encode a sample from the posterior $q(\rvz \mid \vx)$, our method relies on a shared sequence of random samples from the prior $\vz_1, \vz_2, \dots \sim p(\rvz)$, which in practice is realised using a pseudo-random number generator with a shared random seed. The algorithm selects an element $\vz_i$ from the random sequence with high density under the posterior. Then, the code for the sample is simply `$i$', its index in the sequence. Given $i$, the receiver can recover $\vz_i$ by selecting the $i$th element from the shared random sequence. We show that the codelength of `$i$' is close to the relative entropy $\KL{q(\rvz \mid \vx)}{p(\rvz)}$.

\par
Apart from eliminating the requirement for auxiliary bits, REC offers a few further advantages. First, an issue concerning virtually every deep image compression algorithm is that they require a quantized latent space for encoding. This introduces an inherently non-differentiable step to training, which hinders performance and, in some cases, prevents model scaling \citep{hoogeboom2019integer}. Since our method relies on a shared sequence of prior samples, it is not necessary to quantize the latent space and it can be applied to off-the-shelf VAE architectures with continuous latent spaces. To our knowledge, our method is the first image compression algorithm that can operate in a continuous latent space.

\par
Second, since the codelength scales with the relative entropy but not the number of latent dimensions, the method is unaffected by the pruned dimensions of the VAE, i.e. dimensions where the posterior collapses back onto the prior \citep{yang2020variable, lucas2019understanding} (this advantage is shared with bits-back methods, but not others in general).

\par
Third, our method elegantly extends to lossy compression. If we choose to only encode a sample from the posterior using REC without the residuals, the receiver can use the decoder to reconstruct an image that is close to the original. In this setting, the VAE is optimized using an objective that allows the model to maintain high \textit{perceptual quality} even at low bit rates. The compression rate of the model can be precisely controlled during training using a $\beta$-VAE-like training objective \citep{higgins2017beta}. Our empirical results confirm that our REC algorithm is competitive with the state-of-the-art in lossy image compression on the Kodak dataset \citep{kodakdataset}.

\par
The key contributions of this paper are as follows:
\begin{itemize}
  \item iREC, a relative entropy coding method that can encode an image with codelength close to the negative ELBO for VAEs. Unlike prior bits-back methods, it does not require auxiliary bits, and hence it is efficient for encoding single images. We empirically confirm these findings on the Cifar10, ImageNet32 and Kodak datasets.
  \item Our algorithm forgoes the quantization of latent representations entirely, hence it is directly applicable to off-the-shelf VAE architectures with continuous latent spaces.
  \item The algorithm can be applied to lossy compression, where it is competitive with the state-of-the-art on the Kodak dataset.
\end{itemize}

\section{Learned Image Compression}

The goal of compression is to communicate some information between the sender and the receiver using as little bandwidth as possible. Lossless compression methods assign some \textit{code} $C(x)$ to some input $x$, consisting of a sequence of bits, such that the original $x$ can be always recovered from $C(x)$. The efficiency of the method is determined by the average length of $C(x)$. 

All compression methods operate on the same underlying principle: \textit{Commonly occurring patterns are assigned shorter codelengths while rare patterns have longer codelengths}.
This principle was first formalized by \citet{shannon1948mathematical}. Given an underlying distribution $P(\rx)$,\footnote{In this paper use roman letters (e.g. $\rvx$) for random variables and italicized letters (e.g. $\vx$) for their realizations. Bold letters denote vectors. By a slight abuse of notation, we denote $p(\rvx = \vx)$ by $p(\vx)$.}
 where $\rx$ is the input taking values in some set $\X$, the rate of compression cannot be better than $H[P]=\sum_{x \in \X} -P(x)\log P(x)$, the Shannon-entropy of $P$. This theoretical limit is achieved when the codelength of a given $x$ is close to the negative log-likelihood $|C(x)|\approx -\log P(x)$. Methods that get close to this limit are referred to as entropy coding methods, the most prominent ones being Huffman coding and arithmetic coding \citep{huffman1952method, witten1987arithmetic}.

The main challenge in image compression is that the distribution $P(\rx)$ is not readily available. Methods either have to hand-craft $P(\rx)$ or learn it from data. The appeal in using generative models to learn $P(\rx)$ is that they can give significantly better approximations than traditional approaches.

\subsection{Image Compression Using Variational Autoencoders}

A Variational Autoencoder (VAE, \citet{kingma2013auto}) is a generative model that learns the underlying distribution of a dataset in an unsupervised manner. It consists of a pair of neural networks called the \textit{encoder} and \textit{decoder}, that are approximate inverses of each other. The encoder network takes an input $\vx$ and maps it to a posterior distribution over the latent representations $q_\phi(\rvz \mid \vx)$. The decoder network maps a latent representation $\vz \sim q_\phi(\rvz \mid \vx)$ to the conditional distribution $P_\theta(\rvx \mid \vz)$. Here, $\phi$ and $\theta$ denote the parameters of the encoder and the decoder, respectively. These two networks are trained jointly by maximizing a lower bound to the marginal log-likelihood $\log P(\vx)$, the Evidence Lower BOund $\mathcal{L}(\vx, \phi, \theta)$ (ELBO):
\begin{equation}
\log P(\vx) \geq \mathcal{L}(\vx, \phi, \theta) = \underbrace{\Exp{\log P_\theta(\vx \mid \rvz)}{\rvz \sim q_\phi(\rvz \mid \vx)}}_{\text{conditional log-likelihood}} - \underbrace{\KL{q_\phi(\rvz \mid \vx)}{p(\rvz)}}_{\text{relative entropy}} \,.
\end{equation}

\begin{figure}
     \centering
     \begin{subfigure}[b]{0.37\textwidth}
         \centering
         \includegraphics[width=\textwidth]{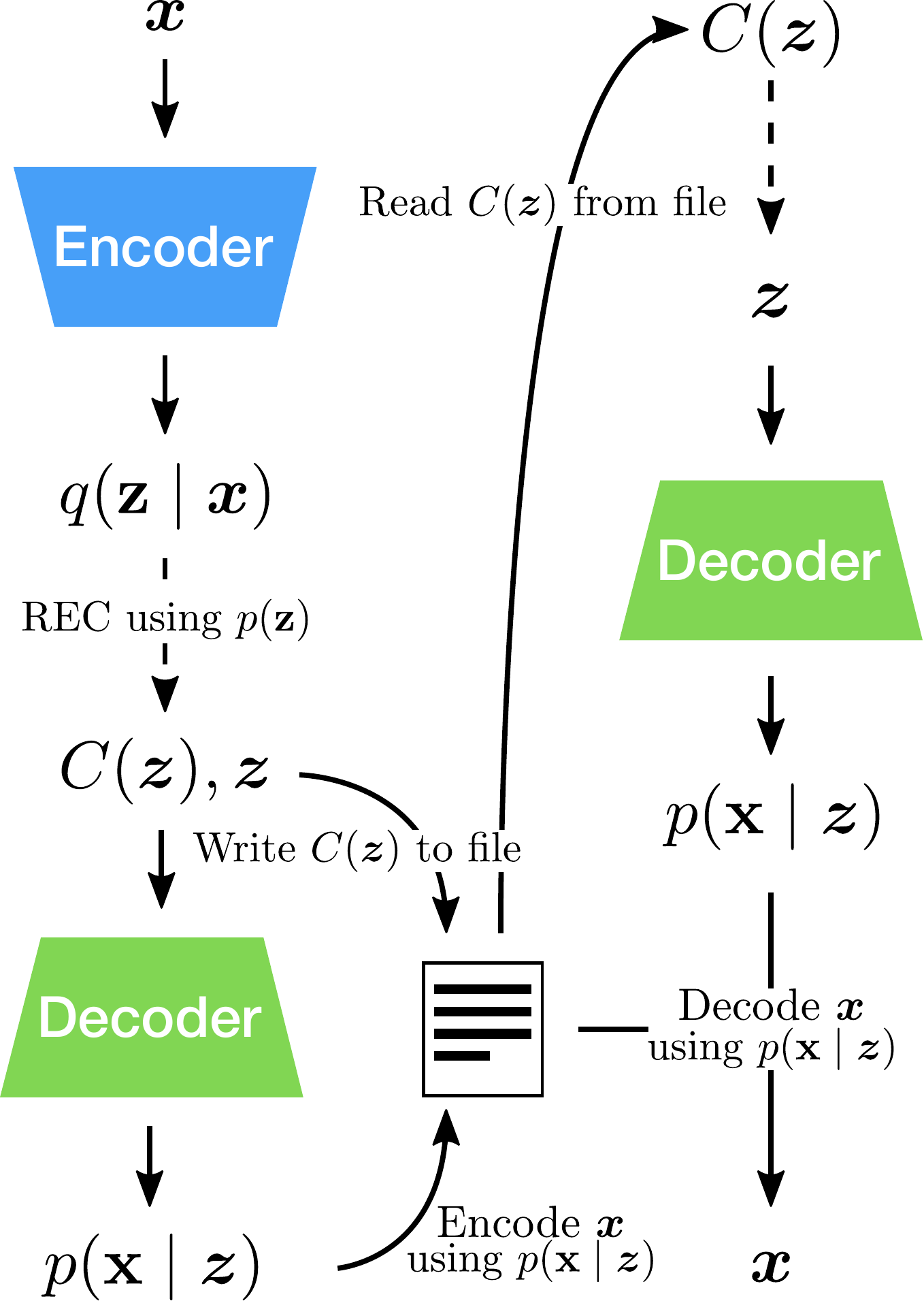}
         \caption{}
         \label{fig:rec_lossless_compression}
     \end{subfigure}
     \hfill
     \begin{subfigure}[b]{0.37\textwidth}
         \centering
         \includegraphics[width=\textwidth]{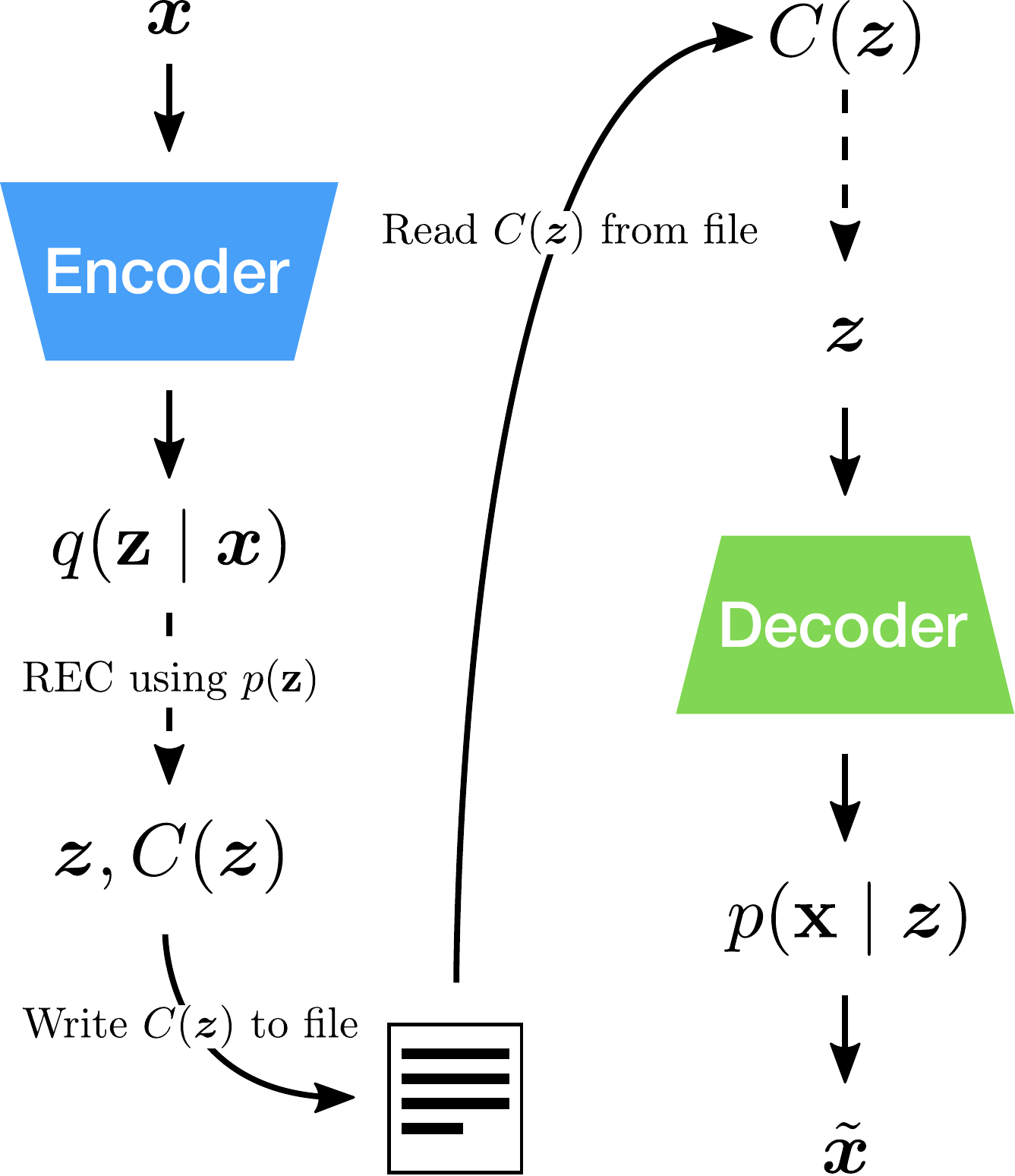}
         \caption{} 
         \label{fig:rec_lossy_compression}
     \end{subfigure}
     \begin{subfigure}[b]{0.24\textwidth}
         \includegraphics[width=\textwidth]{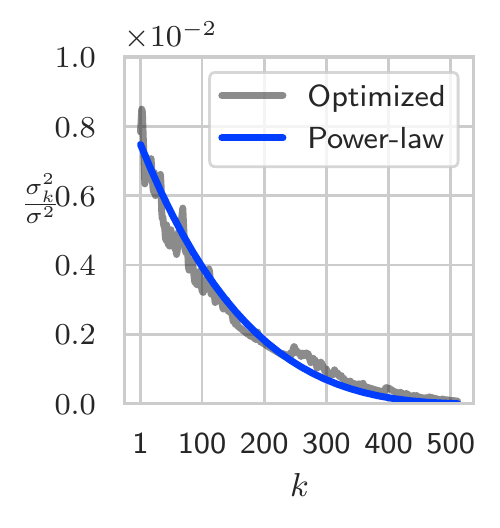} 
         \caption{}
         \label{fig:aux_kl_power_law}
         \includegraphics[width=\textwidth]{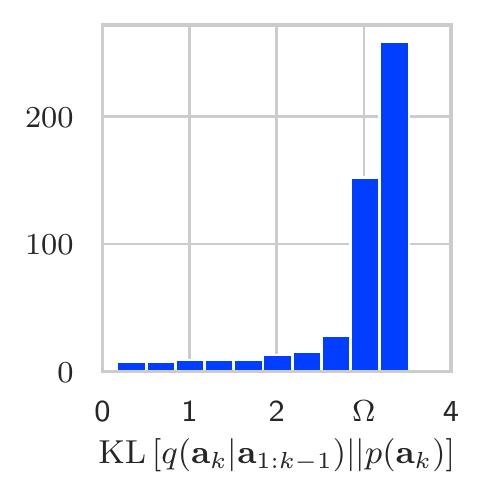}
         \caption{}
         \label{fig:aux_kl_dist}
     \end{subfigure}
     \vspace{-0.1cm}
        \caption{\textbf{(a)} Lossless compression using REC \textbf{(b)} Lossy compression using REC \textbf{(c)} The variances of the coding distributions of the auxiliary variables. We observe that the individually optimized values are well approximated by a power-law. \textbf{(d)} The relative entropies of the auxiliary variables are near or below $\Omega$. ((c) and (d) depict statistics from the 23rd stochastic layer of a 24-layer ResNet VAE, since this layer contains the majority of the model's total relative entropy.)}
        \label{fig:rec_compression}
\end{figure}

The VAE can be used to realize a non-linear form of transform coding \citep{goyal2001theoretical} to perform image compression. Given an image $\vx$, the sender to maps it to its latent posterior $q(\rvz \mid \vx)$, and communicates a sample $\vz \sim q(\rvz \mid \vx)$ to the receiver (where we omit $\phi$ and $\theta$ for notational ease). Our proposed algorithm can accomplish this with communication cost close to the relative entropy $\KLqp$. Given $\vz$, the receiver can use the decoder to  obtain the conditional distribution $P(\rvx \mid \vz)$, which can be used for both lossless and lossy compression. Figure \ref{fig:rec_compression} depicts both processes.

\par
For \textbf{lossless} compression, the sender can use an entropy coding method with $P(\rvx \mid \vz)$ to encode the residuals. The cost of communicating the residuals is the negative log-likelihood $-\log P(\rvx \mid \vz)$ which yields a combined codelength close to the negative ELBO. (See Figure \ref{fig:rec_lossless_compression}.)

\par
For \textbf{lossy} compression, the most commonly used approach is to take the mean of the conditional distribution to be the approximate reconstruction $\tilde{\vx} = \Exp{\rvx}{\rvx \sim {P(\rvx \mid \vz)}}$. This yields a reconstruction close to the original image, while only having to communicate $\vz$. (See Figure \ref{fig:rec_lossy_compression}.)

The remaining question is how to communicate a sample $\vz$ from the posterior $q(\rvz \mid \vx)$ given the shared prior $p(\rvz)$. The most widely given answer is the \textbf{quantization} of the latent representations,
which are then encoded using entropy coding \citep{theis2017lossy, balle2016end}.
This approach is simple to use but has two key weaknesses. First, because of the quantized latent space, the posterior is a discrete probability distribution, which is significantly more difficult to train with gradient descent than its continuous counterpart. It requires the use of gradient estimators and does not scale well with depth \citep{hoogeboom2019integer}. A second known shortcoming of this method is that the codelength scales with the number of latent dimensions even when those dimensions as pruned by the VAE, i.e. the posterior coincides with the prior and the dimension, therefore, carries no information \citep{yang2020variable, lucas2019understanding}. 

An alternative to quantization in lossless compression is \textbf{bits-back coding} \citep{townsend2019practical}. It uses a string of auxiliary bits to encode $\vz$, which is then followed by encoding the residuals. When compressing multiple images, bits-back coding reuses the code of already compressed images as auxiliary bits to compress the remaining ones, bringing the asymptotic cost close to the negative ELBO. However, due to the use of auxiliary bits, it is inefficient to use for single images. 

Relative Entropy Coding (REC), rectifies the aforementioned shortcomings. We propose a REC algorithm that can encode $\vz$ with codelength close to the relative entropy, without requiring quantization or auxiliary bits. It is effective for both lossy and lossless compression.

\section{Relative Entropy Coding}
\par
Relative Entropy Coding (REC) is a \textit{lossless compression paradigm} that solves the problem of communicating a sample from the posterior distribution $q(\rvz \mid \rvx)$ given the shared prior distribution $p(\rvz)$. In more general terms, the sender wants to communicate a sample $\vz \sim q(\rvz)$ to the receiver from a target distribution (the target distribution is only known to the sender) with a coding distribution $p(\rvz)$ shared between the sender and the receiver, given a \textit{shared source of randomness}. That is, over many runs,  the empirical distribution of the transmitted $\vz$s converges to $q(\rvz)$. Hence, REC is a \textit{stochastic} coding scheme, in contrast with entropy coding which is fully deterministic. 

We refer to algorithms that achieve communication cost provably close to  $\KL{q(\rvz)}{p(\rvz)}$ as REC algorithms. We emphasize the counter-intuitive notion, that communicating a stochastic sample from $q(\rvz)$ can be much cheaper than communicating any \textit{specific} sample $\vz$. 
For example, consider the case when $p(\rvz) = q(\rvz)=\Norm{\rvz \mid 0, I}$. First, consider the naive approach of sampling $\vz \sim q(\rvz)$ and encoding it with  entropy coding. The expected codelength of $\vz$ is $\infty$ since the Shannon entropy of a Gaussian random variable is $\infty$. Now consider a REC approach: the sender could simply indicate to the receiver that they should draw a sample from the shared coding distribution $p(\rvz)$, which has $\Oh(1)$ communication cost.
This problem is studied formally in \citet{harsha2007communication}. They show that the relative entropy is a lower bound to the codelength under mild assumptions. As part of a formal proof, they present a rejection sampling algorithm that in practice is computationally intractable even for small problems, however, it provides a basis for our REC algorithm presented below.

\subsection{Relative Entropy Coding with Index Coding}
\par
We present Index Coding (iREC), a REC algorithm that scales to the needs of modern image compression problems.
The core idea for the algorithm is to rely on a shared source of randomness between the sender and the receiver, which takes the form of an infinite sequence of random samples from the prior $\vz_1, \vz_2, \dots \sim p(\rvz)$. This can be practically realized using a pseudo-random number generator with a shared random seed. Then, to communicate an element $\vz_i$ from the sequence, it is sufficient to transmit its index, i.e. $C(\vz_i)=i$. From $i$ the receiver can reconstruct $\vz_i$ by using the shared source of randomness to generate $\vz_1, \vz_2, \dots$ and then selecting the $i$th element.

iREC is based on the importance sampling procedure proposed in \citet{havasi2018minimal}. Let $M = \lceil\exp\left(\KL{q(\rvz)}{p(\rvz)}\right)\rceil$. Our algorithm draws $M$ samples $\vz_1, \hdots, \vz_M \sim p(\rvz)$, then selects $\vz_i$ with probability proportional to the importance weights $P_{\text{accept}}(\vz_m) \propto \frac{q(\vz_m)}{p(\vz_m)}$ for $m \in \{1,\hdots M\}$. Although $\vz_i$ is not an unbiased sample from $q(\rvz)$, \citet{havasi2018minimal} show that considering $M$ samples is sufficient to ensure that the bias remains low. The cost of communicating $z_i$ is simply $\log M\approx \KL{q(\rvz)}{p(\rvz)}$, since we only need to communicate $i$ where $1 \leq i \leq M$.

Importance sampling has promising theoretical properties, however, it is still infeasible to use in practice because $M$ grows exponentially with the relative entropy. To drastically reduce this cost, we propose sampling a sequence of auxiliary variables instead of sampling $q(\rvz)$ directly.

\subsubsection{Using Auxiliary Variables}
\label{sec:using_aux_vars}
We propose breaking $\rvz$ up into a sequence of $K$ auxiliary random variables $\rva_{1:K} = \rva_1, \dots, \rva_K$ with independent coding distributions $p(\rva_1) \dots , p(\rva_K)$ such that they fully determine $\rvz$, i.e. $\rvz = f(\rva_{1:K})$ for some function $f$.
Our goal is to derive target distributions $q(\rva_k\mid\rva_{1:k-1})$ for each of the auxiliary variables given the previous ones, such that by sampling each of them via importance sampling, i.e. $\va_k\sim q(\rva_k\mid\rva_{1:k-1})$ for $k \in \{1,\hdots K\}$, we get a sample $\vz = f(\va_{1:K}) \sim q(\rvz)$.
This implies the auxiliary coding distributions $p(\rva_k)$ and target distributions $q(\rva_k \mid \rva_{1:k-1})$ must satisfy the marginalization properties
\begin{equation}
\label{eq:aux_marginalization_cond}
    p(\rvz) = \int \delta(f(\va_{1:K}) - \rvz)p(\va_{1:K})\mathrm{d}\va_{1:K} \quad \text{and} \quad q(\rvz) = \int \delta(f(\va_{1:K}) - \rvz)q(\va_{1:K})\mathrm{d}\va_{1:K}\,,
\end{equation}
where $\delta$ is the Dirac delta function,
$p(\va_{1:K})=\prod_{k=1}^Kp(\va_k)$ and $q(\va_{1:K})=\prod_{k=1}^K q(\va_k\mid \rva_{1:k-1})$. Note that the coding distributions $p(\rva_1) \dots , p(\rva_K)$ and $f$ can be freely chosen subject to Eq \ref{eq:aux_marginalization_cond}.

\par
The cost of encoding each auxiliary variable using importance sampling is equal to the relative entropy between their corresponding target and coding distributions. Hence, to avoid introducing an overhead to the overall codelength, the targets must satisfy $\,\, \KL{q(\rva_{1:k})}{p(\rva_{1:k})}=\KL{q(\rvz)}{p(\rvz)}$.\footnotemark
\footnotetext{We use here the definition $\KL{q(\rx \mid \ry)}{p(\rx)} = \Exp{\log \frac{q(\rx \mid \ry)}{p(\rx)} }{x, y \sim p(\rx, \ry)}$  \citep{cover2012elements}.}
To ensure this, for fixed $f$ and joint auxiliary coding distribution $p(\rva_{1:K})$, the joint auxiliary target distributions must have the form
\begin{equation}
\label{eq:aux_posterior_defn}
    q(\rva_{1:k}):=\int p(\rva_{1:k} \mid \vz)q(\vz)\,\mathrm{d}\vz  \quad \text{for }k \in \{1 \hdots K\} \,.
\end{equation}
These are the only possible auxiliary targets that satisfy the condition on the sum of the relative entropies, which we formally show in the supplementary material.

\begin{figure}[t]
  \resizebox{\columnwidth}{!}{
  \centering\setlength\tabcolsep{1.25em}
  \begin{tabular}{cc}
    \multirow{2}{0.575\textwidth}[\dimexpr1.08in-\baselineskip-0.5\abovecaptionskip-0.8ex\relax]{\subcaptionbox{\label{alg:irec_encoder}}{
    \setlength{\interspacetitleruled}{-.4pt}%
                \begin{algorithm}[H]
                 \KwData{$q(\rvz)$}
                 \KwResult{$(i_1, \dots, i_K)$}
                 $K \gets \left\lceil \frac{\KL{q(\rvz)}{p(\rvz)}}{\Omega} \right \rceil$ \\
                  $M \gets \left\lceil \exp\left(\Omega (1 + \epsilon)\right) \right\rceil$ \\
                 $S_0\gets \{()\}$ \\
                 \For{$k\leftarrow 1$ \KwTo $K$}{
                  $\va_{k_1}, \dots, \va_{k_{M}} \widesim{R} p(\va_k) $  \\
                  \vspace{0.1cm}
                  $\hat{S}_k \gets S_{k-1} \times \{1, \hdots, M\}$ \\
                  $S_k \gets \underset{(j_1, \hdots,\, j_k) \in \hat{S}_{k}}{\argtop_{B}} \left \{ \frac{q\left( \va_{1_{j_1}}, \hdots,\, \va_{k_{j_k}} \right)}{p\left( \va_{1_{j_1}},\hdots, \va_{k_{j_k}} \right)} \right \}$ \\
                 }
                 $(i_1, \hdots, i_K) \gets \underset{(j_1, \hdots, j_K) \in S_K}{\argmax} \left \{ \frac{q \left( f\left(\va_{1_{j_1}},\dots, \va_{K_{j_K}}\right) \right) }{p\left( f\left(\va_{1_{j_1}},\dots, \va_{K_{j_K}}\right) \right)} \right \}$
                \end{algorithm}
                \vspace{-0.2cm}
    }
    } & \subcaptionbox{\label{alg:irec_decoder}}{
    \setlength{\interspacetitleruled}{-.4pt}%
    \begin{minipage}{.32\linewidth}
            \begin{algorithm}[H]
             \KwData{$(i_1, \dots, i_K)$}
             \KwResult{$\vz$}
              $M \gets \left\lceil \exp\left(\Omega (1 + \epsilon)\right) \right\rceil$ \\
             \For{$k\leftarrow 1$ \KwTo $K$}{
              $\va_{k_1}, \dots, \va_{k_{M}} \widesim{R} p(\rva_k)$ \\
             }
             $\vz \gets f(\va_{1_{i_1}},\dots, \va_{K_{i_K}})$ \\
            \end{algorithm}
            \end{minipage}
            \vspace{-0.2cm}
    } \\
    & \subcaptionbox{\label{fig:beam_search_num_beams}}{
    \hspace{-1cm}\includegraphics[width=0.36\linewidth]{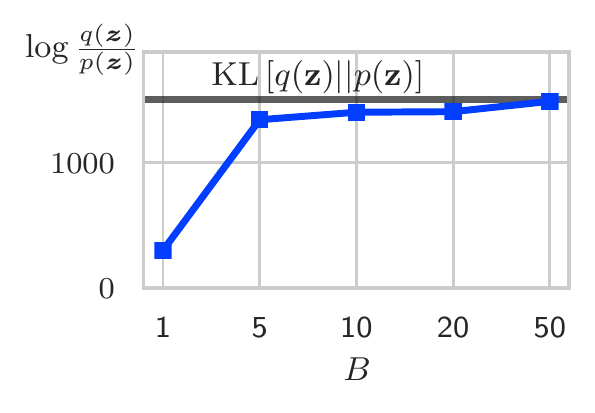}\vspace{-0.4cm}
    }
  \end{tabular}
  }
  \caption{\textbf{(a)} iREC encoder \textbf{(b)} iREC decoder \textbf{(c)} Beam search ensures that $\log \frac{q(\vz)}{p(\vz)}$ is close to the relative entropy. $B$ is the number of beams. Plotted using the 23rd stochastic layer of a 24-layer ResNet VAE, since this layer contains the majority of the model's total relative entropy.
    Here, $\widesim{R}$ indicates sampling using a pseudo-random number generator with random seed $R$, and $\argtop_B$ selects the arguments of the top $B$ ranking elements in a set. }
    \label{fig:beam_search}
\end{figure}

\paragraph{Choosing the forms of the auxiliary variables.}
Factorized Gaussian priors are a popular choice for VAEs. For a Gaussian coding distribution $p(\rvz) = \mathcal{N}(0, \sigma^2 I)$, we propose  $\rvz = \sum_{k = 1}^K\rva_k$,  
$p(\rva_k)=\mathcal{N}(0, \sigma_k^2 I)$ for $k \in \{1,\hdots, K\}$ such that $\sum_{k=1}^K\sigma^2_k=\sigma^2\,$. The targets $q(\rva_k\mid \rva_{1:k - 1})$ in this case turn out to be Gaussian as well, and their form is derived in the supplementary material.

To guarantee that every auxiliary variable can be encoded via importance sampling, the relative entropies should be similar across the auxiliary variables, i.e. $\KL{q(\rva_k\mid \rva_{1:k-1})}{p(\rva_k)} \approx \Omega$,
where $\Omega$ is a hyperparameter (we used $\Omega=3$ in our experiments). 
This yields $K = \left\lceil \KL{q(\rvz)}{p(\rvz)} /\, \Omega \right\rceil$ auxiliary variables in total.
We initially set the auxiliary coding distributions by optimizing their variances $\sigma_k^2$ on a small validation set to achieve relative entropies close to $\Omega$. Later, we found that the ratio of the variance $\sigma^2_k$ of the $k$-th auxiliary variable to the remaining variance $\sigma^2 - \sum_{j = 1}^{k - 1}\sigma_j^2$
is well approximated by the power law $(K + 1 - k)^{-0.79}$, as shown in Figure \ref{fig:aux_kl_power_law}. In practice, we used this approximation to set each $\sigma^2_k$. With these auxiliary variables, we fulfil the requirement of keeping the individual relative entropies near or below $\Omega$ as shown empirically in Figure \ref{fig:aux_kl_dist}. To account for the auxiliary variables whose relative entropy slightly exceeds $\Omega$, in practice, we draw $M = \left\lceil \exp\left(\Omega (1 + \epsilon)\right) \right\rceil$ samples, where $\epsilon$ is a small non-negative constant (we used $\epsilon=0.2$ for lossless and $\epsilon=0.0$ for lossy compression), leading to a $(1+\epsilon)$  increase of the codelength.

\subsubsection{Reducing the Bias with Beam Search}
\par
An issue with naively applying the auxiliary variable scheme is that the cumulative bias of importance sampling each auxiliary variable adversely impacts the compression performance, potentially leading to higher distortion and hence longer codelength. 
\par
To reduce this bias, we propose using a beam search algorithm (shown in Figure \ref{fig:beam_search}) to search over multiple possible assignments of the auxiliary variables. We maintain a set of the $B$ lowest bias samples from the auxiliary variables and used the log-importance weight $\log \frac{q(\vz)}{p(\vz)}$ as a heuristic measurement of the bias. For an unbiased sample $\log \frac{q(\vz)}{p(\vz)} \approx \KL{q(\rvz)}{p(\rvz)}$, but for a biased sample,  $\log \frac{q(\vz)}{p(\vz)} \ll \KL{q(\rvz)}{p(\rvz)}$. For each auxiliary variable $\rva_k$ in the sequence, we combine the $B$ lowest bias samples  for $\rva_{1:k-1}$ with the $M$ possible importance samples for $\rva_k$. To choose the $B$ lowest bias samples for the next iteration, we take the top $B$ samples with the highest importance weights $\frac{q(\va_{1:k})}{p(\va_{1:k})}$ out of the $B\times M$ possibilities. At the end, we select $\va_{1:K}$ with the highest importance weight $\frac{q \left( f\left(\va_{1:K}\right) \right) }{p\left( f\left(\va_{1:K}\right) \right)}=\frac{q \left( \vz \right)}{p\left( \vz\right)}$. 

\subsubsection{Determining the hyperparameters}
\par
Finding good values for $\Omega, \epsilon$ and $B$ is crucial for the good performance of our method. We want as short a codelength as possible, while also minimizing computational cost. Therefore, we ran a grid search over a reasonable range of parameter settings for the lossless compression of a small number of ImageNet32 images. We compare the efficiency of settings by measuring the codelength overhead they produced in comparison to the ELBO, which represents optimal performance. We find that for reasonable settings of $\Omega$ (between 5-3) and for fixed $\epsilon$, regular importance sampling ($B=1$) gives between 25-80\% overhead, whereas beam search with $B=5$ gives 15-25\%, and with $B=20$ it gives 10-15\%. Setting $B>20$ does not result in significant improvements in overhead, while the computational cost is heavily increased. Thus, we find that 10-20 beams are sufficient to significantly reduce the bias as shown in Figure \ref{fig:beam_search_num_beams}. The details of our experimental setup and the complete report of our findings can be found in the supplementary material.
\section{Experiments}
\label{sec:experiments}
\par
We compare our method against state-of-the-art lossless and lossy compression methods. Our experiments are implemented in \texttt{TensorFlow} \citep{tensorflow2015-whitepaper} and are publicly available at \url{https://github.com/gergely-flamich/relative-entropy-coding}.

\subsection{Lossless Compression}
\label{subsec:lossless_compression_experiments}

\par
We compare our method on single image lossless compression (shown in Table \ref{tab:lossless_results}) against PNG, WebP and FLIF, Integer Discrete-Flows \citep{hoogeboom2019integer} and the prominent bits-back approaches: Local Bits-Back Coding \citep{ho2019compression}, BitSwap \citep{kingma2019bit} and HiLLoC \citep{townsend2020hilloc}. 

\par
Our model for these experiments is a ResNet VAE (RVAE) \citep{kingma2016improved} with 24 Gaussian stochastic levels. This model utilizes skip-connections to prevent the posteriors on the higher stochastic levels to collapse onto the prior and achieves an ELBO that is competitive with the current state-of-the-art auto-regressive models. For better comparison, we used the exact model used by \citet{townsend2020hilloc}\footnote{We used the publicly available trained weights published by the authors of \citet{townsend2020hilloc}.} trained on ImageNet32. The three hyperparameters of iREC are set to $\Omega = 3$, $\epsilon = 0.2$ and $B = 20$.
Further details on our hyperparameter tuning process are included in the supplementary material.

\par
We evaluated the methods on Cifar10 and ImageNet32 comprised of $32\times 32$ images, and the Kodak dataset comprised of full-sized images. For Cifar10 and ImageNet32, we used a subsampled test set of size 1000 due to the speed limitation of our method (currently, it takes $\sim 1$ minute to compress a $32\times32$ image and 1-10 minutes to compress a large image).
By contrast, decoding with our method is fast since it does not require running the beam search procedure.
\par
iREC significantly outperforms other bits-back methods on all datasets since it does not require auxiliary bits, although it is still slightly behind non-bits-back methods as it has a $\sim 20\%$ overhead compared to the ELBO due to using $\epsilon=0.2$.

\newcolumntype{d}[1]{D{.}{.}{#1}}
\newcommand\mc[1]{\multicolumn{1}{c}{#1}} 
\newcommand*{\B}[1]{\ifmmode\bm{#1}\else\textbf{#1}\fi}

\begin{table}[t]
\centering
\caption{Single image, lossless compression performance in bits per dimension (lower is better). The best performing bits-back or REC method is highlighted for each dataset. The asymptotic rates are included in parenthesis where they are different from the single image case. To calculate the number of bits needed for single images, we added the number of auxiliary bits required to the asymptotic compression rate as reported in the respective papers.}
\resizebox{\columnwidth}{!}{
\begingroup
\setlength{\tabcolsep}{10pt} 
\renewcommand{\arraystretch}{1.5}
\label{tab:lossless_results}
\begin{tabular}{ll*{3}{d{3.3}}}
\toprule
  &            & \mc{Cifar10 (32x32)}  & \mc{ImageNet32 (32x32)} & \mc{Kodak (768x512)} \\ 
    \midrule
\multirow{4}{*}{\textit{Non bits-back}} & PNG        &  5.87 &  6.39        &   4.35   \\
  & WebP       & 4.61  & 5.29 &   3.20    \\
  & FLIF       & 4.19  & 4.52  & 2.90  \\ 
   & IDF        & 3.34  & 4.18  & -         \\ \midrule
\multirow{3}{*}{\textit{Bits-back}\footnotemark }    & LBB        & 54.96\ (3.12) & 55.72\ (3.88) & -         \\ 
  & BitSwap   & 6.53\  (3.82) & 6.97\ (4.50) & -     \\
  & HiLLoC     & 24.51\ (3.56) & 26.80\ (4.20) & 17.5\ (3.00)     \\ \midrule
\multirow{2}{*}{\textit{REC}}  & iREC (Ours) & \bm{4}.\bm{18}  &  \bm{4}.\bm{91} & \bm{3}.\bm{67}    \\
  & ELBO (RVAE)  &  [3.55] & [4.18] & [3.00] \\
\bottomrule
\end{tabular}
\endgroup
}
\end{table}

\subsection{Lossy Compression}
\label{subsec:lossy_compression_experiments}
\par

\footnotetext{To overcome the issue of the inefficiency of bits-back methods for single or small-batch image compression, in practice an efficient single-image compression method (e.g.\ FLIF) is used to encode the first few images and only once the overhead of using bits-back methods becomes negligible do we switch to using them (see e.g.\ \citet{townsend2020hilloc}).}

\begin{figure}[t]
     \centering
    \resizebox{0.882\columnwidth}{!}{
     \hfill
     \begin{subfigure}[t]{0.45\textwidth}
         \centering
         \includegraphics[width=\textwidth]{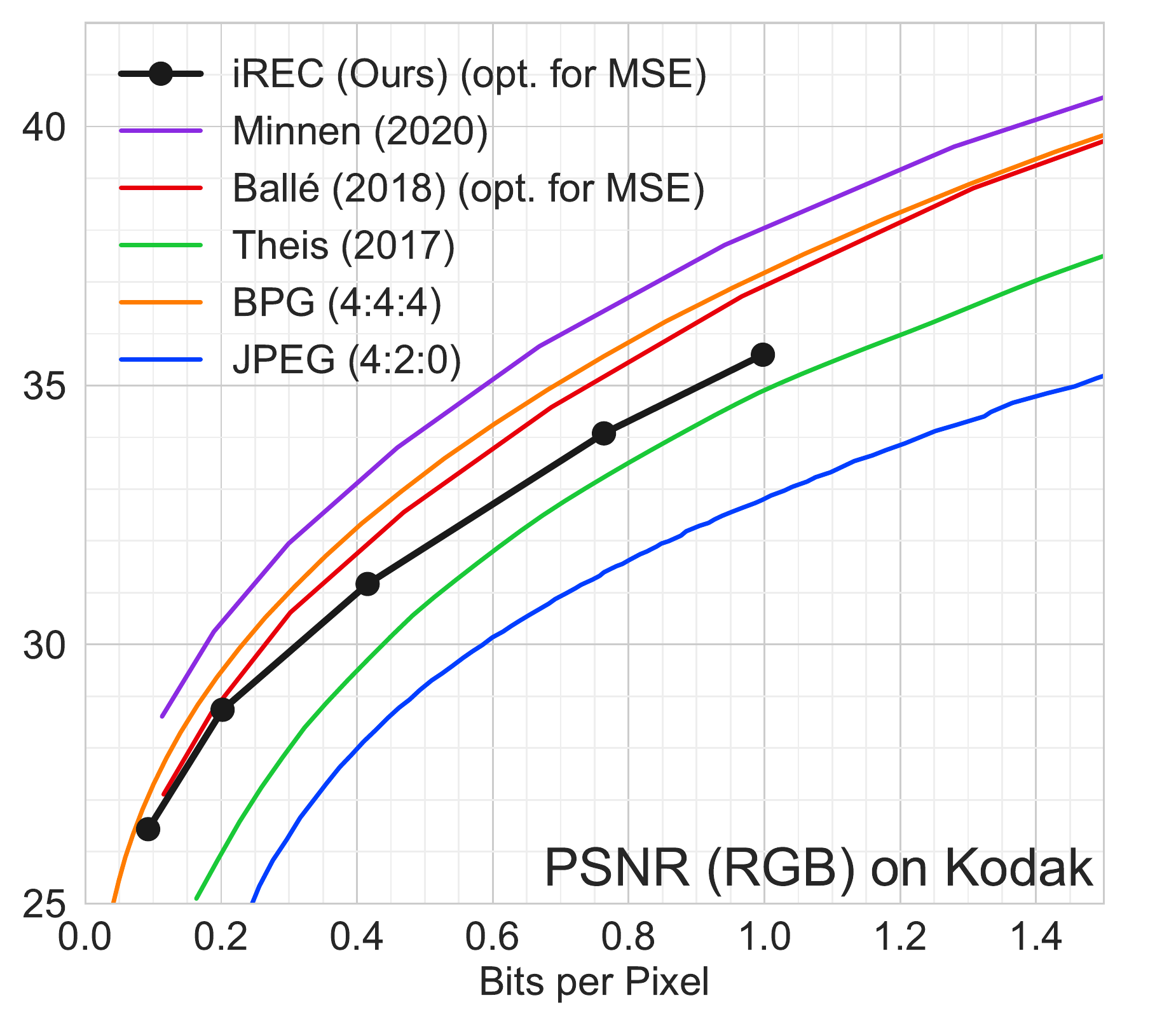}
         \vspace{-0.7cm}
         \caption{}
         \vspace{-0.1cm}
         \label{fig:method_psnr_comparison}
     \end{subfigure}
     \hfill
     \begin{subfigure}[t]{0.45\textwidth}
         \centering
         \includegraphics[width=\textwidth]{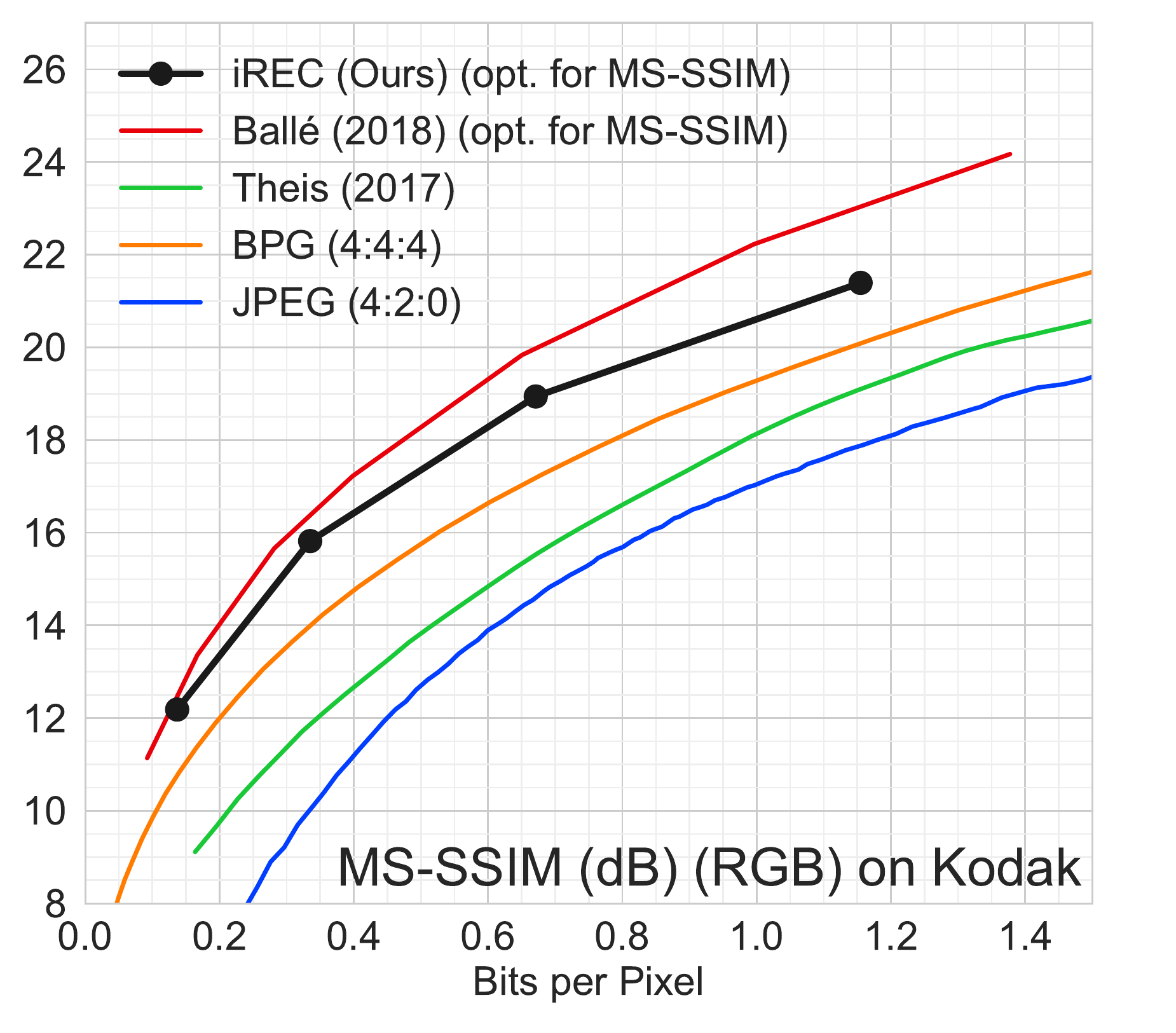}
         \vspace{-0.7cm}
         \caption{}
         \vspace{-0.1cm}
         \label{fig:method_ms_ssim_comparison}
     \end{subfigure}
     \hfill
     }
        \caption{Comparison of REC against classical methods such as JPEG, BPG and competing ML-based methods. \textbf{(a)} PSNR comparisons \textbf{(b)} MS-SSIM comparisons in decibels, calculated using the formula $ -10 \log_{10} (1 - \text{MS-SSIM})$. See the supplementary material for more comparisons.}
        \label{fig:method_comparison}
\end{figure}

On the lossy compression task, we present average rate-distortion curves calculated using the PSNR \citep{psnr} and MS-SSIM \citep{wang2004image} quality metrics on the Kodak dataset, shown in Figure \ref{fig:method_comparison}. On both metrics, we compare against JPEG, BPG, \citet{theis2017lossy} and \citet{balle2018variational}, with the first two being classical methods and the latter two being ML-based. Additionally, on PSNR we compare against \citet{minnen2020channel}, whose work represents the current state-of-the-art to the best of our knowledge.\footnote{For all competing methods, we used publicly available data at \url{https://github.com/tensorflow/compression/tree/master/results/image_compression}.}

\par
We used the architecture presented in \citet{balle2018variational} with the latent distributions changed to Gaussians and a few small modifications to accommodate this; see the supplementary material for precise details. Following \citet{balle2018variational}, we trained several models using $\mathcal{L}_\lambda(\vx, \phi, \theta) = \lambda D(\vx, \hat{\vx}) - \KL{q_\phi(\rvz \mid \vx)}{p(\rvz)}$, where $\hat{\vx}$ is the reconstruction of the image $\vx$, and $D(\cdot, \cdot) \in \{\text{MSE}, \text{MS-SSIM}\}$ is a differentiable distortion metric.\footnote{In practice we used $1 - \text{MS-SSIM}$ as the loss function with power factors set to $\alpha=\beta=\gamma = 1$.} Varying $\lambda$ in the loss yields models with different rate-distortion trade-offs. We optimized 5 models for MSE with $\lambda \in \{0.001, 0.003, 0.01, 0.03, 0.05\}$ and 4 models for MS-SSIM with $\lambda \in \{0.003, 0.01, 0.03, 0.08\}$. The hyperparameters of iREC were set this time to $\Omega = 3$, $\epsilon = 0$ and $B = 10$. As can be seen in Figure \ref{fig:method_comparison}, iREC is competitive with the state-of-the-art lossy compression methods on both metrics.

\par

\section{Related Work}
\label{sec:related_work}

\par
Multiple recent bits-back papers build on the work of \citet{townsend2019practical}, who first proposed BB-ANS. These works elevate the idea from a theoretical argument to a practically applicable algorithm. \citet{kingma2019bit} propose BitSwap, an improvement to BB-ANS that allows it to be applied to hierarchical VAEs. \citet{ho2019compression} extend the work to general flow models, which significantly improves the asymptotic compression rate. Finally,  \citet{townsend2020hilloc} apply the original BB-ANS algorithm to full-sized images and improve its run-time by vectorizing its implementation.

\par
 All previous VAE-based lossy compression methods use entropy coding with quantized latent representations. They propose different approaches to circumvent the non-differentiability that this presents during training. Most prominently, \citet{balle2016end} describe a continuous relaxation of quantization based on dithering. Building on this idea, \citet{balle2018variational} introduce a 2-level hierarchical architecture, and \citet{minnen2018joint, lee2019context, minnen2020channel} explore more expressive latent representations, such as using learned adaptive context models to aid their entropy coder, and autoregressive latent distributions.

\section{Conclusion}
\label{sec:discussion}
\par
This paper presents iREC, a REC algorithm, which extends the importance sampler proposed by \citet{havasi2018minimal}. It enables the use of latent variable models (and VAEs in particular) with continuous probability distributions for both lossless and lossy compression. 

\par
Our method significantly outperforms bits-back methods on lossless single image compression benchmarks and is competitive with the asymptotic performance of competing methods. On the lossy compression benchmarks, our method is competitive with the state-of-the-art for both the PSNR and MS-SSIM perceptual quality metrics. Currently, the main practical limitation of bits-back methods, including our method, is the compression speed, which we hope to improve in future works.


\section{Acknowledgements}

Marton Havasi is funded by EPSRC. We would like to thank Stratis Markou and Ellen Jiang for the helpful comments on the manuscript.

\section{Broader Impact}
\par
Our work presents a novel data compression framework and hence inherits both its up and downsides. In terms of positive societal impacts, data compression reduces the bandwidth requirements for many applications and websites, making them more inexpensive to access. This increases accessibility to online content in rural areas with limited connectivity or underdeveloped infrastructure. Moreover, it reduces the energy requirement and hence the environmental impact of information processing systems. However, care must be taken when storing information in a compressed form for long time periods, and backwards-compatibility of decoders must be maintained, as data may otherwise be irrevocably lost, leading to what has been termed the Digital Dark Ages \citep{digitaldarkage}.


\bibliography{references}
\bibliographystyle{neurips_2020_conference}
\newpage
\appendix
\section{Deriving the posterior distributions for the auxiliary variables}
\par
In this section, we derive Equation (3) presented in the main text, and show how this general result can be applied to the case where $\rvz$ and all $\rva_k$s are selected to be Gaussian, and $f(\rva_{1:K}) = \sum_{k = 1}^K \rva_k$, and show a simple scheme to implement the proposed auxiliary variable method in practice.
\subsection{Deriving the $q(\rva_{1:k})$s - general case}
\label{sec:deriving_aux_targets}
Here we show that the form of the auxiliary posterior $q(\rva_{1:k})$ presented in Equation (3) is the only suitable choice such that the KL divergence remains unchanged. Concretely, fix $f$ and the auxiliary coding distributions $p(\rva_{k}\mid \rva_{1:k - 1})$ for $k \in \{1, \hdots, K\}$. Now, we want to find $q(\rva_{1:K})$ such that the condition
\begin{equation}
\label{eq:aux_kl_sum_cond_appendix}
    \KL{q(\rva_{1:K})}{p(\rva_{1:K})}=\KL{q(\rvz)}{p(\rvz)}
\end{equation}
is satisfied.
Observe that
\begin{equation}
\begin{aligned}
\label{eq:joint_kl_two_ways}
    \KL{q(\rvz, \rva_{1:K})}{p(\rvz, \rva_{1:K})} &= \KL{q(\rvz)}{p(\rvz)} + \KL{q(\rva_{1:K} \mid \rvz)}{p(\rva_{1:K} \mid \rvz)} \\
    &=\KL{q(\rva_{1:K})}{p(\rva_{1:K})} + \KL{q(\rvz \mid  \rva_{1:K})}{p(\rvz \mid \rva_{1:K})},
\end{aligned}
\end{equation}
where the two equalities follow from breaking up the joint KL using the chain rule of relative entropies in two different ways. Notice, that since $\rvz = f(\rva_{1:K})$ is a deterministic relationship, $q(\rvz \mid  \rva_{1:K}) = p(\rvz \mid  \rva_{1:K}) = \delta \left(f(\rva_{1:K}) - \rvz \right)$. Hence, we have $\KL{q(\rvz \mid  \rva_{1:K})}{p(\rvz \mid \rva_{1:K})} = 0$. Using this fact to simplify Eq \ref{eq:joint_kl_two_ways}, we get
\begin{equation}
    \label{eq:simplified_joint_kl_decomp}
    \KL{q(\rvz)}{p(\rvz)} + \KL{q(\rva_{1:K} \mid \rvz)}{p(\rva_{1:K} \mid \rvz)} =\KL{q(\rva_{1:K})}{p(\rva_{1:K})}.
\end{equation}
We see that our original condition in Eq \ref{eq:aux_kl_sum_cond_appendix} for the auxiliary targets is satisfied when $\KL{q(\rva_{1:K} \mid \rvz)}{p(\rva_{1:K} \mid \rvz)} = 0$. Applying the chain rule of relative entropies $K$ times, the condition can be rewritten as
\begin{equation}
     \KL{q(\rva_{1:K} \mid \rvz)}{p(\rva_{1:K} \mid \rvz)} = \sum_{k=1}^K \KL{q(\rva_k\mid \rva_{1:k - 1}, \rvz)}{p(\rva_k \mid \rva_{1:k - 1}, \rvz)} = 0.
\end{equation}
Due to the non-negativity of relative entropy, the above is satisfied if and only if $\KL{q(\rva_k\mid \rva_{1:k - 1}, \rvz)}{p(\rva_k \mid \rva_{1:k - 1}, \rvz)} = 0 \,\,\forall\,\, k \in \{1,\hdots, K\}$. This further implies, that 
\begin{equation}
\label{eq:appendix_coding_target_equality}
        q(\rva_k\mid \rva_{1:k - 1}, \rvz) = p(\rva_k \mid \rva_{1:k - 1}, \rvz)\quad\text{as well as}\quad
        q(\rva_{1:k} \mid \rvz) = p(\rva_{1:k} \mid \rvz)
\end{equation}
for all $k \in \{1,\hdots, K\}$. Now, fix $k$. Then, we have
\begin{equation}
\begin{aligned}
    q(\rva_{1:k}) &= \int q(\rva_{1:k} \mid \vz)q(\vz) \mathrm{d}\vz\\
    &= \int p(\rva_{1:k} \mid \vz)q(\vz) \mathrm{d}\vz
\end{aligned}
\end{equation}
by Eq \ref{eq:appendix_coding_target_equality}, as required.
\par
\subsection{An iterative scheme}
\par
A simple way to directly implement an auxiliary variable coding scheme described in Section 3.1.1 is to draw samples sequentially from the conditional auxiliary targets $q(\rva_k \mid \rva_{1:k - 1})$. Here, we note the following pair recursive relationships:
\begin{equation}
\label{eq:appendix_useful_identity_pair}
    \begin{aligned}
        q(\rva_k \mid \rva_{1:k - 1}) &= \int q(\rva_k \mid \rva_{1:k - 1}, \vz)q(\vz \mid \rva_{1:k - 1}) \mathrm{d} \vz \\
        &= \int p(\rva_k \mid \rva_{1:k - 1}, \vz)q(\vz \mid \rva_{1:k - 1}) \mathrm{d} \vz
        &\\\\
        q(\rvz \mid \rva_{1:k}) &= \frac{q(\rvz, \rva_k \mid \rva_{1:k - 1})}{q(\rva_k \mid \rva_{1:k - 1})} = \frac{p(\rva_k \mid \rva_{1:k - 1}, \vz)q(\rvz \mid \rva_{1:k - 1})}{q(\rva_k \mid \rva_{1:k - 1})},
    \end{aligned}
\end{equation}
where the second equality in both identities follows from Eq \ref{eq:appendix_coding_target_equality}. We can therefore proceed to sequentially encode the $\rva_k$s using Algorithm \ref{alg:auxiliary_coding}.
\begin{figure}[t]
  \centering\setlength\tabcolsep{1.25em}
\begin{algorithm}[H]
 \KwData{$q(\rvz), p(\rva_k \mid \rva_{1:k - 1}) \quad\forall k \in \{1,\hdots, K\}$}
 \KwResult{$S = (i_1, \hdots, i_K)$}
 $S \gets ()$\\
 $q_0(\rvz) \gets q(\rvz)$\\
 \For{$k\leftarrow 1$ \KwTo $K$}{
  $q(\rva_k \mid \rva_{1:k - 1}) \gets \int p(\rva_k \mid \rva_{1:k - 1}, \vz)q_{k - 1}(\vz) \mathrm{d} \vz$ \\
  \,\\
  $i_k, \va_k \gets \mathrm{iREC}(q(\rva_k \mid \rva_{1:k - 1}), p(\rva_k \mid \rva_{1:k - 1}))$ \\
  Append $i_k$ to $S$ \\
  \,\\
  $q_k(\rvz) \gets \frac{p(\rva_k \mid \rva_{1:k - 1}, \vz)q_{k - 1}(\rvz)}{q(\rva_k \mid \rva_{1:k - 1})}$
 }
 \caption{Simple sequential auxiliary coding scheme. The $\mathrm{iREC(\cdot, \cdot)}$ function performs index coding as described in Section 3.1.}
\label{alg:auxiliary_coding}
\end{algorithm}
 \end{figure}

\subsection{The independent Gaussian sum case}

\par
In this section we derive the form of the auxiliary coding targets $q(\rva_k \mid \rva_{1:k - 1})$ and the conditionals $q(\rvz \mid \rva_k)$ in the case when $\rva_k$ are all independent and Gaussian distributed, and $\rvz = f(\rva_{1:K}) = \sum_{k = 1}^K \rva_k$. We only derive the result in the univariate case, extending to the diagonal covariance case is straight forward. However, to keep notation consistent with the rest of the paper, only in this section we will keep using the boldface symbols to denote the random quantities and their realizations.
\par
First, let $p(\rva_k) = \Norm{\rva_k \mid \mu_k, \sigma_k^2}$. Now, define $\rvb_k = \sum_{i = 1}^k \rva_i$, $m_k = \sum_{i = k + 1}^K \mu_i$ and $s_k^2 = \sum_{i = k + 1}^K \sigma_i^2$. Then, since $\rvz = \sum_{k = 1}^K \rva_k$, using the formula for the conditional distribution of sums of Gaussian random variables\footnote{For Gaussian random variables $X,Y$ with means $\mu_x,\mu_y$ and variances $\sigma_x^2,\sigma^2_y$ and $Z=X+Y$, $p(x|z)$ is normally distributed with mean $\mu_x + (z-\mu_x-\mu_y)\frac{\sigma_x^2}{\sigma_x^2 + \sigma_y^2}$ and variance $\frac{\sigma_x^2\sigma_y^2}{\sigma_x^2 + \sigma_y^2}$}, we get
\begin{equation}
    p(\rva_k \mid \rva_{1:k - 1}, \rvz) = \Norm{\rva_k \Biggm\vert \mu_k + (\rvz - \rvb_{k - 1} - m_{k - 1})\frac{\sigma_k^2}{s_{k - 1}^2}, \frac{s_k^2 \sigma_k^2}{s_{k - 1}^2}}.
\end{equation}
Assume that we have already calculated $q(\rvz \mid \rva_{1:k - 1}) = \Norm{\rvz \mid \nu_{k - 1}, \rho_{k - 1}^2}$. From here, we notice that by Eq \ref{eq:appendix_useful_identity_pair}, both quantities of interest are products and integrals of Gaussian densities, and hence after some algebraic manipulation, we get
\begin{equation}
    q(\rva_k \mid \rva_{1:k-1}) = \Norm{\rva_k \Biggm\vert \mu_k + (\nu_{k - 1} - \rvb_{k - 1} - m_{k - 1})\frac{\sigma_k^2}{s_{k - 1}^2}, \frac{s_k^2 \sigma_k^2}{s_{k - 1}^2} + \rho_{k - 1}^2\frac{\sigma_k^4}{s_{k - 1}^4} },
\end{equation}
and 
\begin{equation}
    q(\rvz\mid \rva_{1:k}) = \Norm{\rvz \Biggm\vert 
    \frac{(\rva_k - \mu_k) \rho_{k - 1}^2 s_{k - 1}^2 + (\rvb_{k - 1} + m_{k - 1})\sigma_k^2 \rho_{k - 1}^2 + \nu_{k - 1}s_k^2 s_{k - 1}^2
    }{\sigma_k^2 \rho_{k - 1}^2 + s_{k - 1}^2 s_k^2 }, \frac{\rho_{k - 1}^2 s_{k - 1}^2 s_k^2}{\sigma_k^2 \rho_{k - 1}^2 + s_{k - 1}^2 s_k^2 } }.
\end{equation}
\par
Given the above two formulae, both the simple iterative scheme presented in the previous section as well as the beam search algorithm presented in the main text can be readily implemented.

\section{Hyperparameter Experiments for Beam Search}
\par
Our lossless compression approach has three hyperparameters: $\Omega$, $\epsilon$ and $B$. We tuned these on a small validation set of 10 images from ImageNet32 by sweeping them, and measuring the performance. The codelength can get arbitrarly close to the ELBO, but it requires a significant computational cost. We try to find a good balance between the codelength and the computational cost.

\begin{figure}
\vspace{-3cm}
\hspace{-3.5cm}
\includegraphics[width=1.5\textwidth]{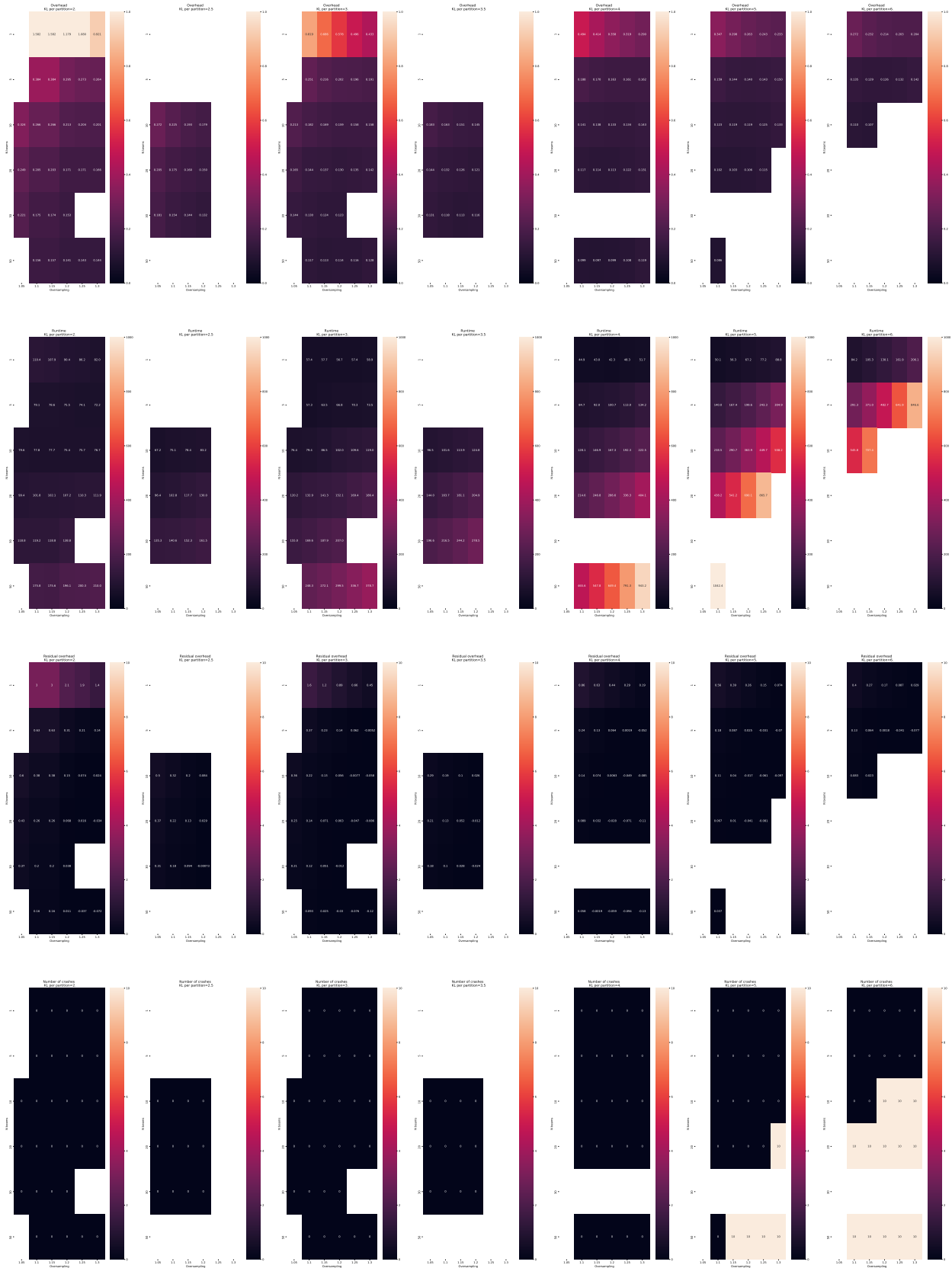}
\vspace{-3.5cm}
\caption{Hyperparameters for lossless compression. $\Omega$ is referred as `KL\_per\_partition', $\epsilon$ is referred as `extra\_samples' and $B$ is referred as `n\_beams'. (On a computer, zoom in to see precise figures)} \label{grid}
\end{figure}

Figure \ref{grid} shows the parameter combinations that we tested. We plot 4 metrics:
\begin{itemize}
    \item 1st row: Overhead in number of bits required compared to the ELBO. A value of 0.2 corresponds to 20\% overhead over the ELBO in codelength.
    \item 2nd row: Time it takes to run the method in seconds.
    \item 3rd row: Residual overhead. The overhead in codelength when only looking at the residual. This helps to estimate the bias in the samples, since if there is no bias, this overhead should be 0.
    \item 4th row: For how many out of the 10 validation images did the method crash. Every crash was cause by memory overflow.
\end{itemize}

\begin{figure}
\includegraphics[width=\textwidth]{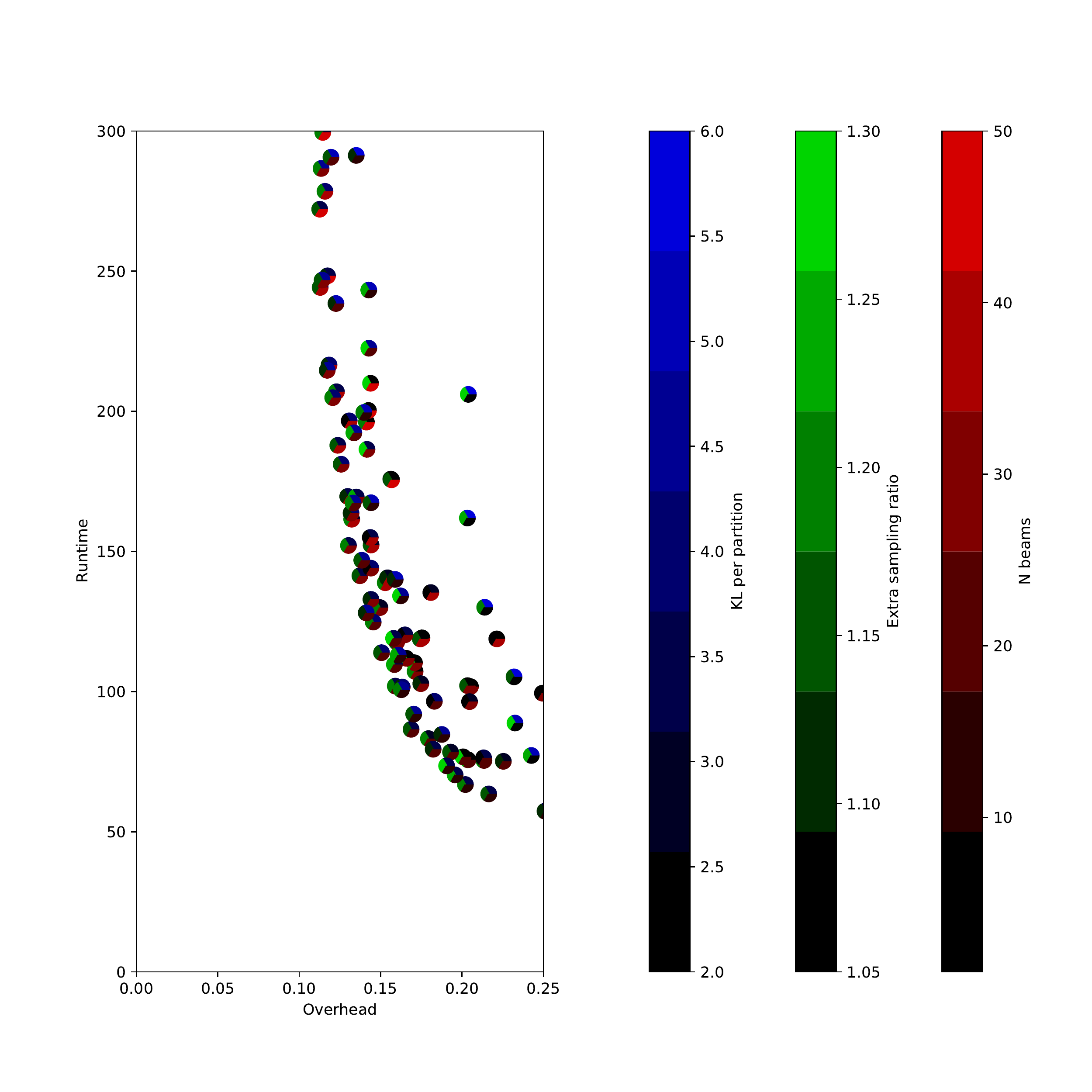}
\caption{Pareto frontier of the hyperparameters.} \label{pareto}
\end{figure}

Figure \ref{pareto} depicts the same data, but plotted in two dimensions: overhead vs time.

\section{Auxiliary variables}
\par
\begin{figure}[htb]
    \centering 
\begin{subfigure}{0.20\textwidth}
  \includegraphics[width=\linewidth]{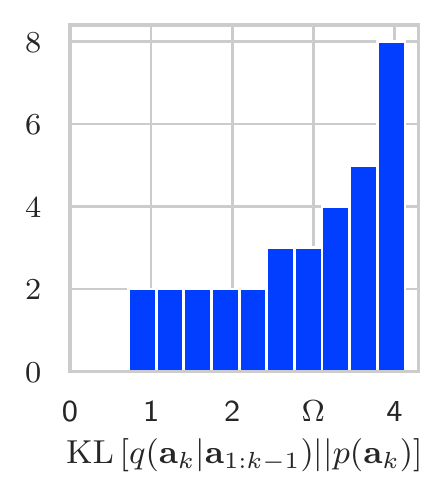}
  \caption{Layer 1}
\end{subfigure}\hfil 
\begin{subfigure}{0.20\textwidth}
  \includegraphics[width=\linewidth]{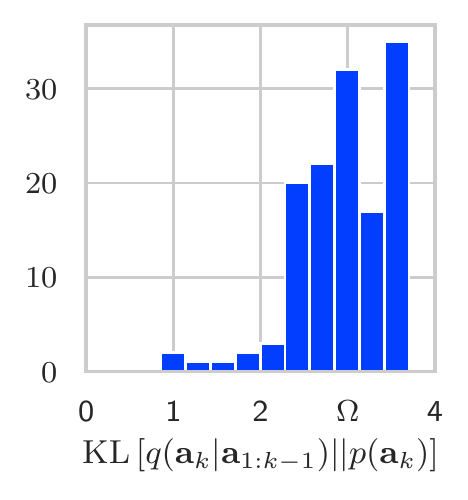}
  \caption{Layer 2}
\end{subfigure}\hfil 
\begin{subfigure}{0.20\textwidth}
  \includegraphics[width=\linewidth]{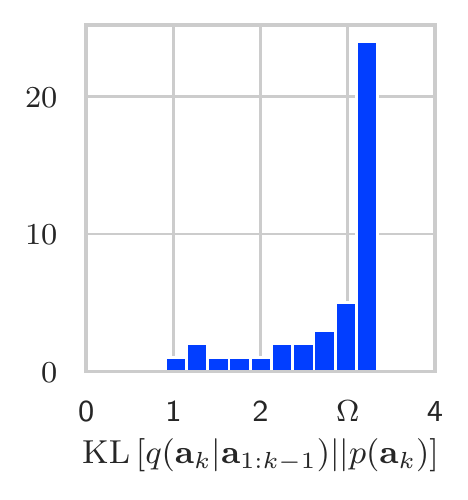}
  \caption{Layer 3}
\end{subfigure}\hfil 
\begin{subfigure}{0.20\textwidth}
  \includegraphics[width=\linewidth]{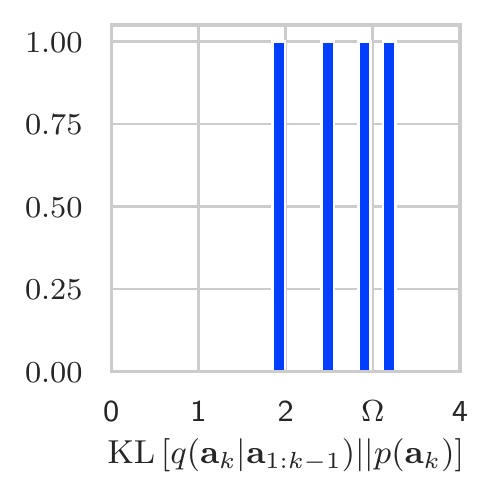}
  \caption{Layer 4}
\end{subfigure}

\medskip

\begin{subfigure}{0.20\textwidth}
  \includegraphics[width=\linewidth]{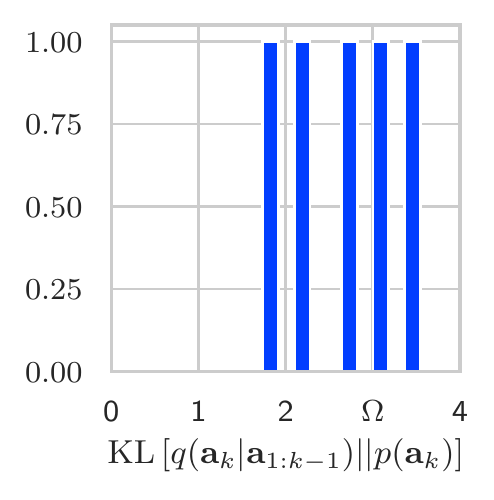}
  \caption{Layer 5}
\end{subfigure}\hfil 
\begin{subfigure}{0.20\textwidth}
  \includegraphics[width=\linewidth]{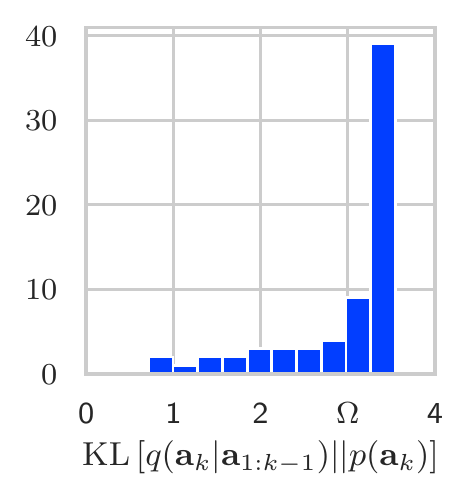}
  \caption{Layer 6}
\end{subfigure}\hfil 
\begin{subfigure}{0.20\textwidth}
  \includegraphics[width=\linewidth]{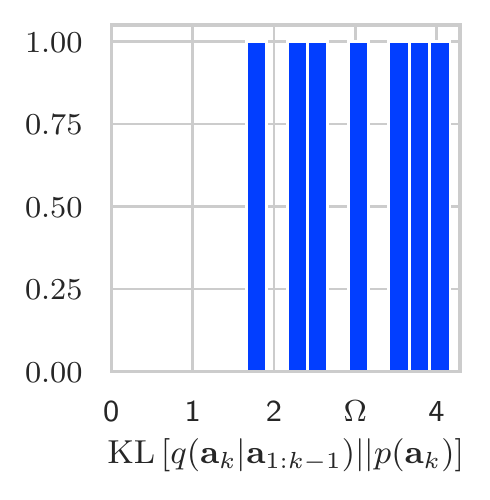}
  \caption{Layer 7}
\end{subfigure}\hfil 
\begin{subfigure}{0.20\textwidth}
  \includegraphics[width=\linewidth]{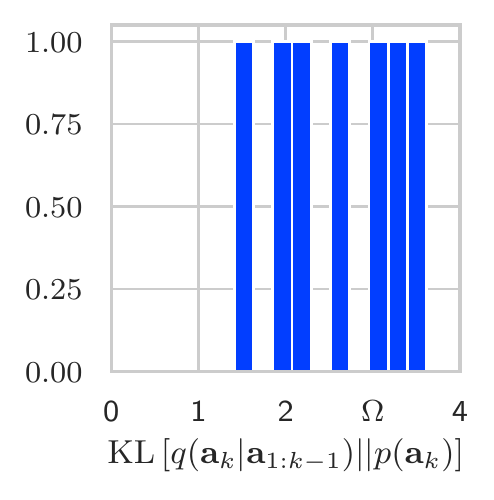}
  \caption{Layer 8}
\end{subfigure}
\medskip

\begin{subfigure}{0.20\textwidth}
  \includegraphics[width=\linewidth]{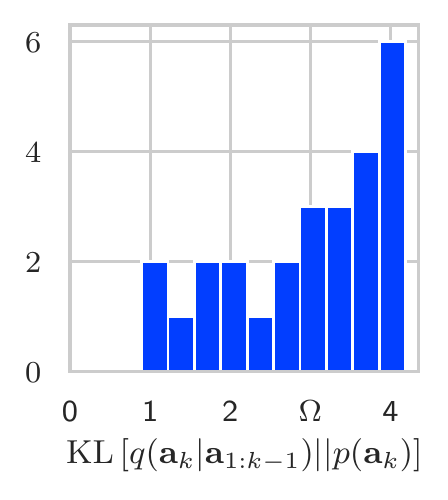}
  \caption{Layer 9}
\end{subfigure}\hfil 
\begin{subfigure}{0.20\textwidth}
  \includegraphics[width=\linewidth]{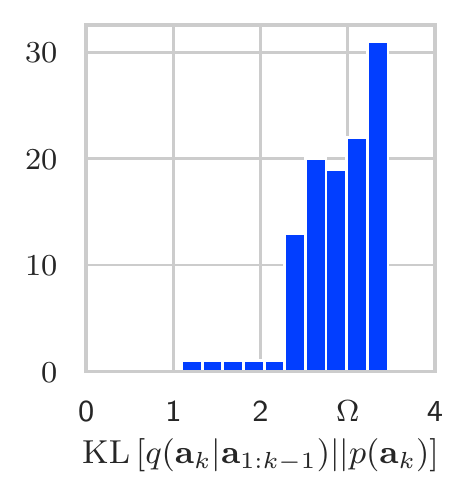}
  \caption{Layer 10}
\end{subfigure}\hfil 
\begin{subfigure}{0.20\textwidth}
  \includegraphics[width=\linewidth]{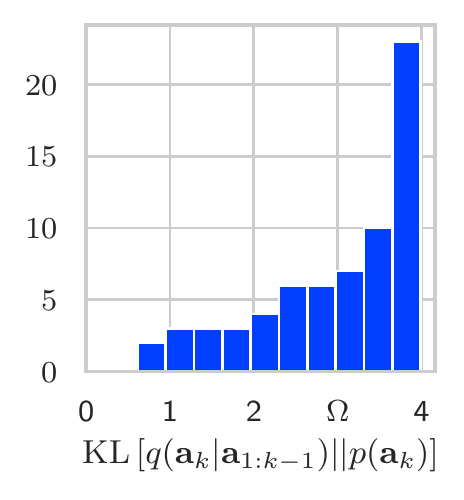}
  \caption{Layer 11}
\end{subfigure}\hfil 
\begin{subfigure}{0.20\textwidth}
  \includegraphics[width=\linewidth]{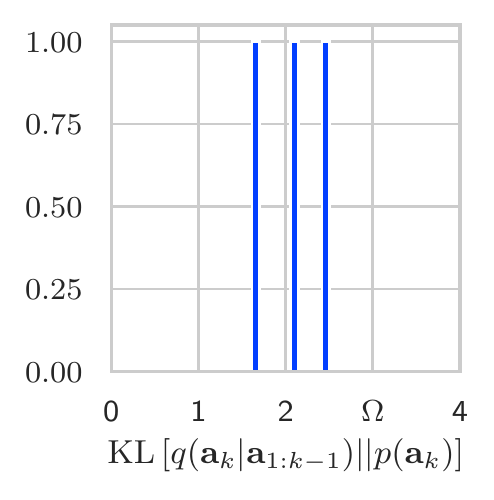}
  \caption{Layer 12}
\end{subfigure}
\medskip

\begin{subfigure}{0.20\textwidth}
  \includegraphics[width=\linewidth]{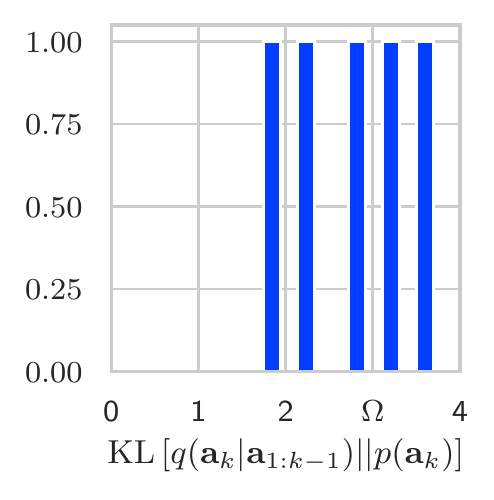}
  \caption{Layer 13}
\end{subfigure}\hfil 
\begin{subfigure}{0.20\textwidth}
  \includegraphics[width=\linewidth]{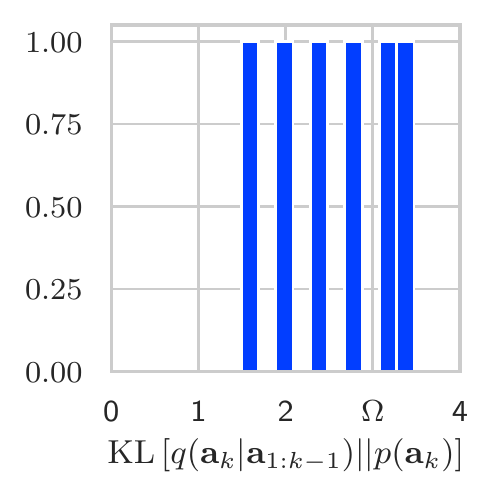}
  \caption{Layer 14}
\end{subfigure}\hfil 
\begin{subfigure}{0.20\textwidth}
  \includegraphics[width=\linewidth]{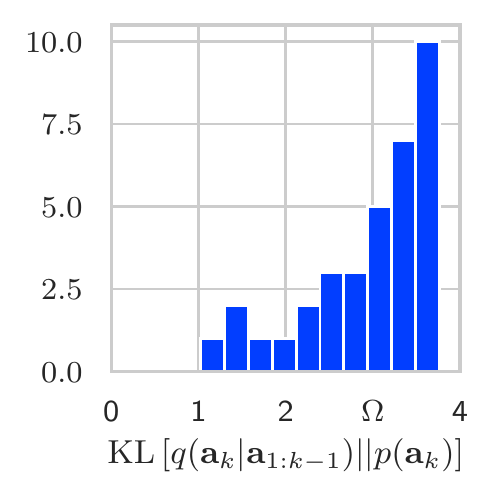}
  \caption{Layer 15}
\end{subfigure}\hfil 
\begin{subfigure}{0.20\textwidth}
  \includegraphics[width=\linewidth]{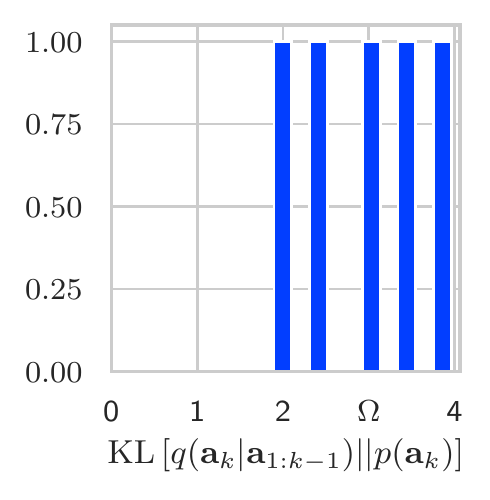}
  \caption{Layer 16}
\end{subfigure}
\medskip

\begin{subfigure}{0.20\textwidth}
  \includegraphics[width=\linewidth]{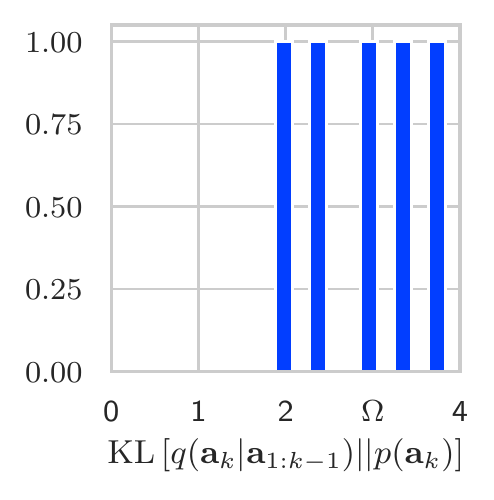}
  \caption{Layer 17}
\end{subfigure}\hfil 
\begin{subfigure}{0.20\textwidth}
  \includegraphics[width=\linewidth]{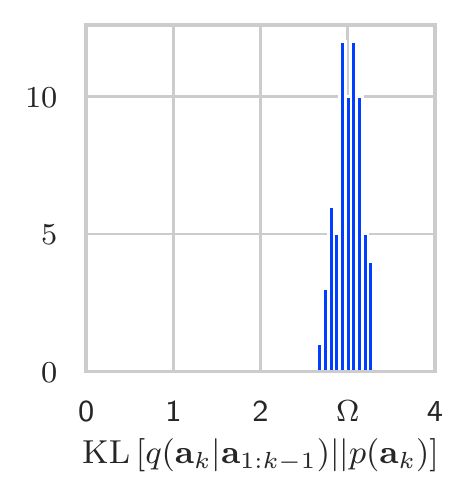}
  \caption{Layer 18}
\end{subfigure}\hfil 
\begin{subfigure}{0.20\textwidth}
  \includegraphics[width=\linewidth]{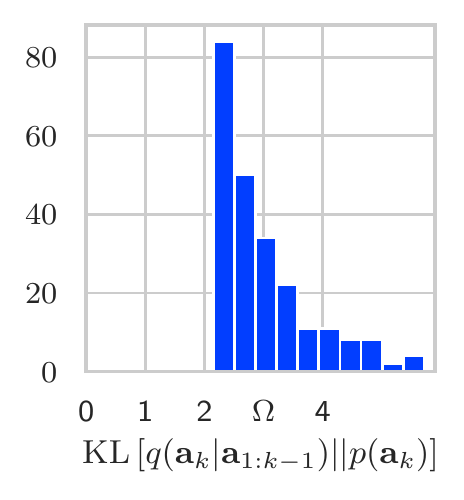}
  \caption{Layer 19}
\end{subfigure}\hfil 
\begin{subfigure}{0.20\textwidth}
  \includegraphics[width=\linewidth]{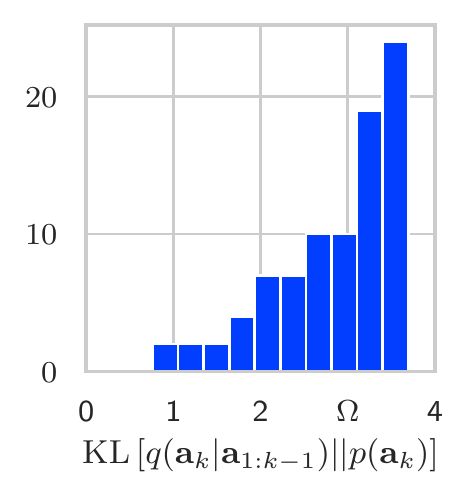}
  \caption{Layer 20}
\end{subfigure}
\medskip

\begin{subfigure}{0.20\textwidth}
  \includegraphics[width=\linewidth]{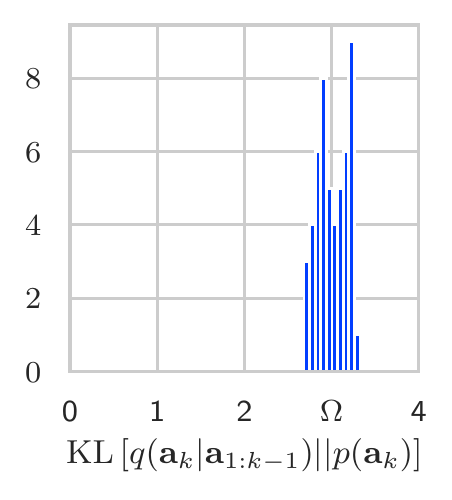}
  \caption{Layer 21}
\end{subfigure}\hfil 
\begin{subfigure}{0.20\textwidth}
  \includegraphics[width=\linewidth]{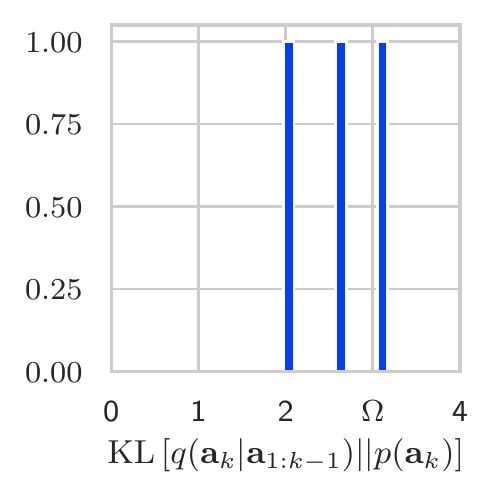}
  \caption{Layer 22}
\end{subfigure}\hfil 
\begin{subfigure}{0.20\textwidth}
  \includegraphics[width=\linewidth]{img/appendix/kl_hist_22.pdf}
  \caption{Layer 23}
\end{subfigure}\hfil 
\begin{subfigure}{0.20\textwidth}
  \includegraphics[width=\linewidth]{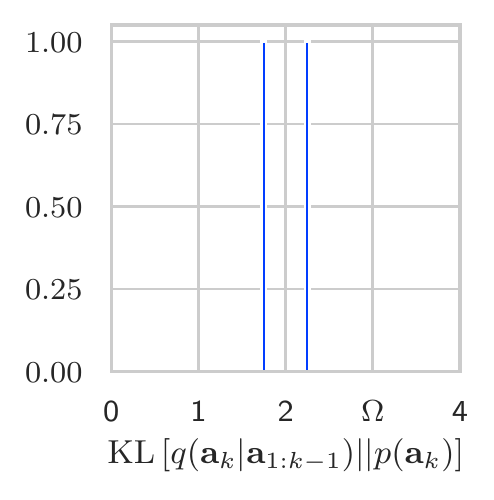}
  \caption{Layer 24}
\end{subfigure}
\caption{Histograms of the relative entropies of the auxiliary variables in a 24 layer RVAE.}
\label{histograms}
\end{figure}

We plotted the 23rd layer of the RVAE in the main paper to demonstrate how the auxiliary variables are able to break up the latent variables such that their relative entropies are close to $\Omega$. Here we include all 24 layers (Figure \ref{histograms}).

\section{Additional Information on the Setup of Lossy Experiments}
\par
\begin{figure}
  \centering
  \includegraphics[width=\textwidth]{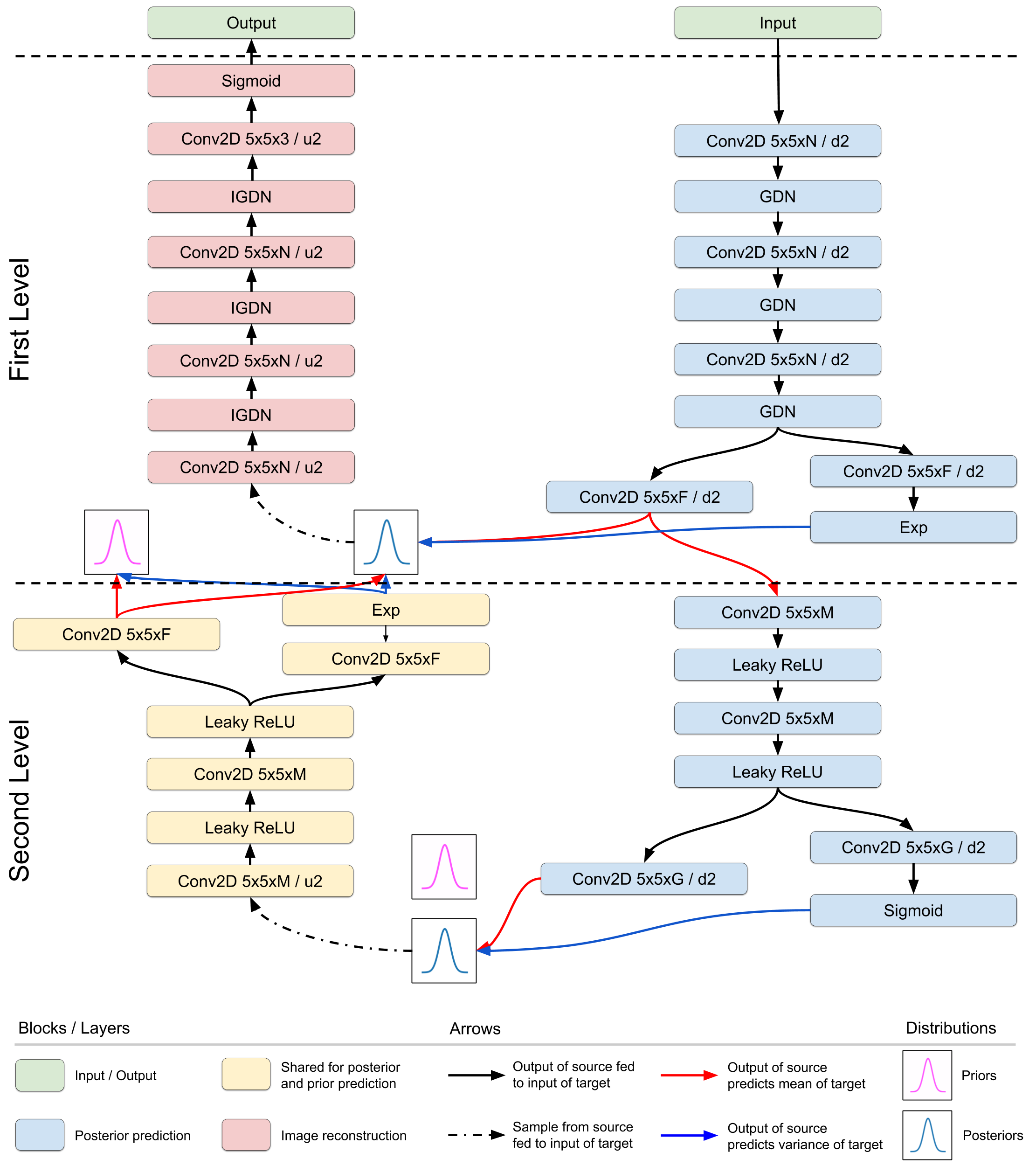}
  \caption[Our Probabilistic Ladder Network (PLNl) architecture.]
  {PLN network architecture. The blocks signal data transformations, the
    arrows signal the flow of information. \textbf{Block descriptions:}
    \textit{Conv2D:} 2D convolutions along the spatial dimensions, where the
    $W\times H \times C / S$ implies a $W \times H$ convolution kernel, with $C$
  target channels and $S$ gives the downsampling rate (given a preceding letter
  ``d'') or the upsampling rate (given a preceding letter ``u''). If the slash
  is missing, it means that there is no up/downsampling. All convolutions operate
  in \texttt{same} mode with zero-padding. \textit{GDN / IGDN:} these are the
  non-linearities described in Ball\'e et al. (2016a). \textit{Leaky ReLU:}
  elementwise non-linearity defined as $\max\{x, \alpha x\}$, where we set
  $\alpha=0.2$. \textit{Sigmoid:} Elementwise non-linearity defined as
  $\frac{1}{1 + \exp\{-x\}}$. }
  \label{fig:pln_architecture}
\end{figure}

We used an appropriately modified version of the architecture proposed by Ball\'e et al. (2018). Concretely, we swapped all latent distributions for Gaussians, and made the encoder and decoder networks two-headed on the appropriate layers to provide both a mean and log-standard deviation prediction for the latent distributions. For a depiction of our architecture, see Figure \ref{fig:pln_architecture}.

\par
During training, we inferred the parameters of the hyperprior as well (Empirical Bayes). We trained every model on the CLIC 2018 dataset for $2 \times 10^5$ iterations with a batch size of 8 image patches. As done in Ball\'e et al. (2018), the patches were $256 \times 256$ and were randomly cropped from the training images. As the dataset is curated for lossy image compression tasks, we performed no further data preprocessing or augmentation. We found that annealing the KL divergence in the beginning (also known as warm-up) did not yield a significant performance increase.

\section{Additional Lossy Compression Results}
\par
In this section we present some additional results that clarify the current shortcomings of iREC and better illustrate its performance on individual images. 
\par
\begin{figure}[h]
     \centering
     \begin{subfigure}[t]{0.49\textwidth}
         \centering
         \includegraphics[width=\textwidth]{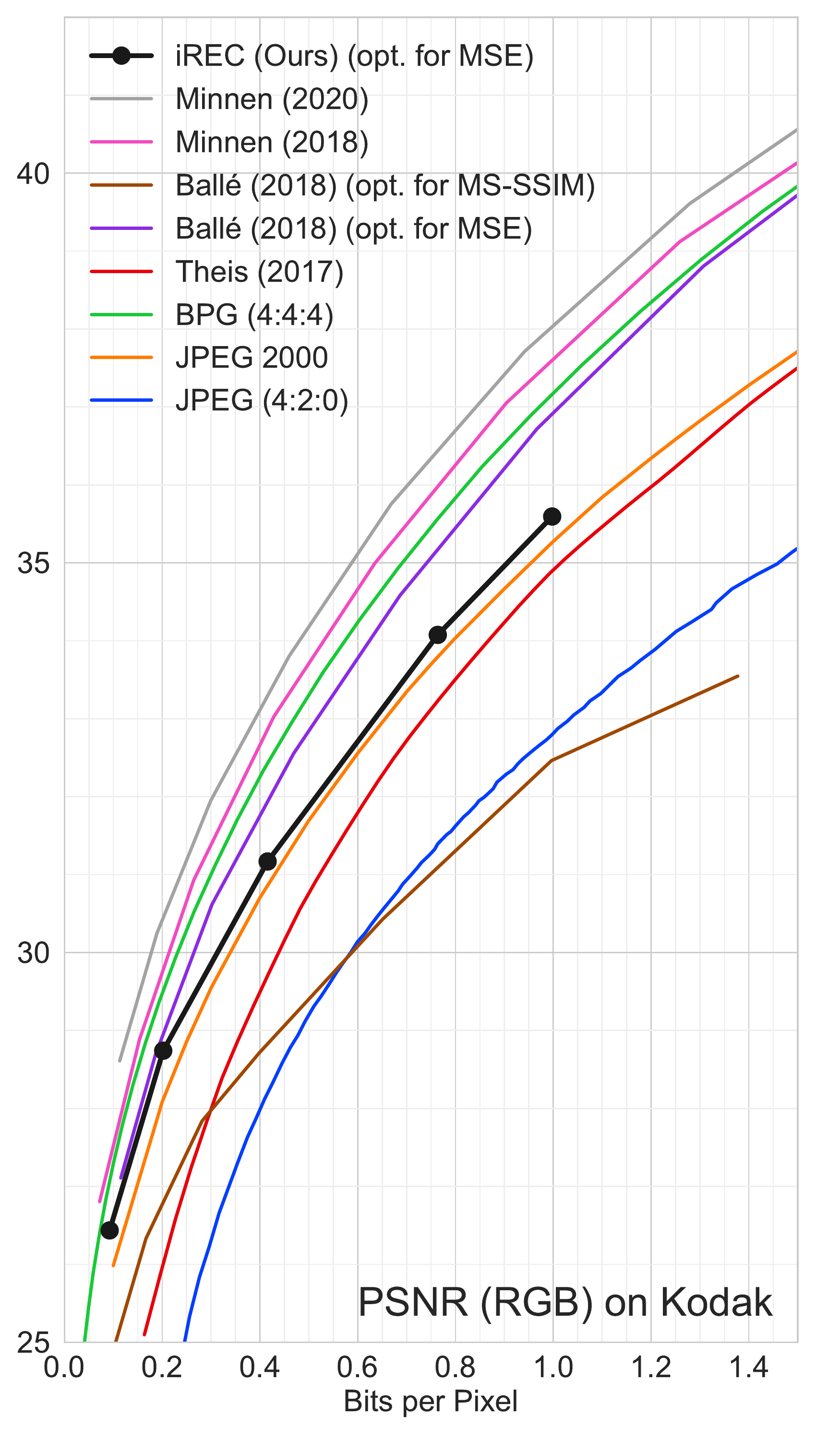}
         \caption{Performance on PSNR.}
         \label{fig:more_psnr}
     \end{subfigure}
     \hfill
     \begin{subfigure}[t]{0.49\textwidth}
         \centering
         \includegraphics[width=\textwidth]{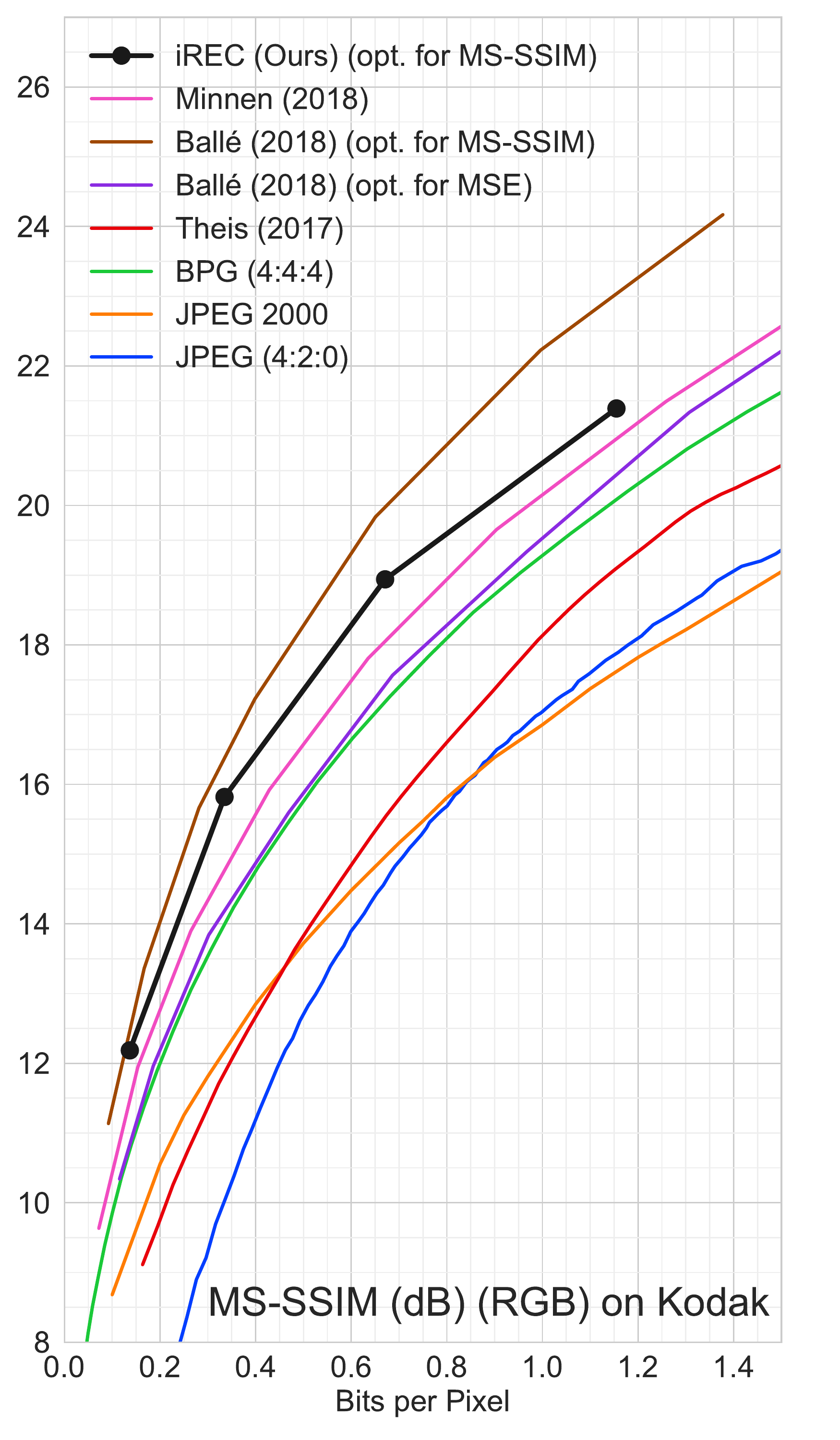}
         \caption{Performance on MS-SSIM.}
         \label{fig:more_ms_ssim}
     \end{subfigure}
        \caption{Comparison of REC against classical methods such as JPEG, JPEG 2000 (OpenJPEG), BPG and competing ML-based methods. MS-SSIM comparisons are in decibels, calculated using the formula $-10 \log_{10} (1 - \text{MS-SSIM})$.}
        \label{fig:more_lossy_comparisons}
\end{figure}

First, we present an extended version of Figure 3 from the main text in Figure \ref{fig:more_lossy_comparisons}.

\subsection{Actual vs Ideal Performance Comparison}
\par
Since iREC is based on importance sampling, the posterior sample it returns will be slightly biased, which affects the distortion of the reconstruction. Furthermore, since it might require setting the oversampling rate $\epsilon > 0$ in some cases, as well as having to communicate some minimal additional side information, the codelength will also be slightly higher than the theoretical lower bound. 
\par
\begin{figure}[h]
     \centering
     \begin{subfigure}[t]{0.49\textwidth}
         \centering
         \includegraphics[width=\textwidth]{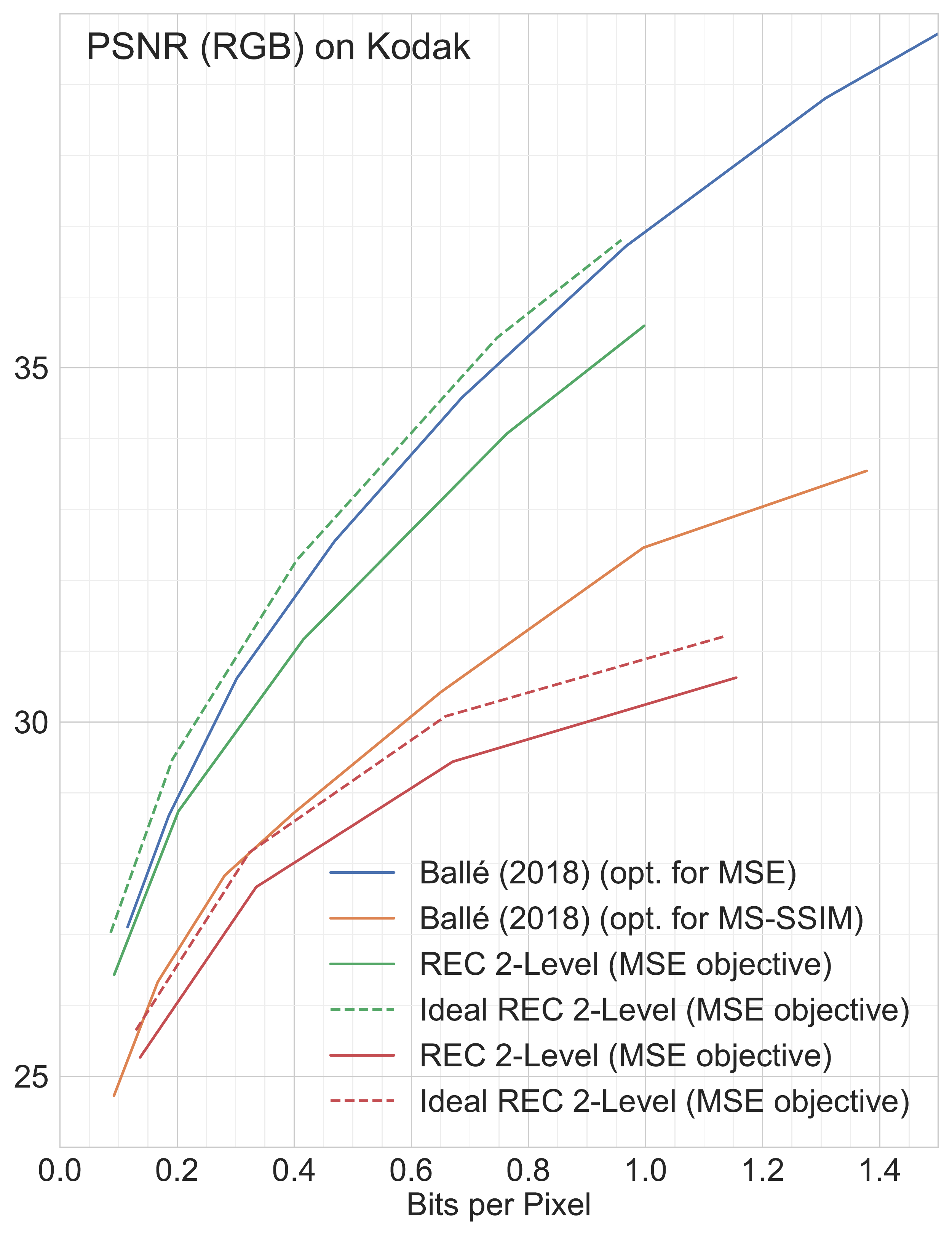}
         \caption{Performance on PSNR.}
         \label{fig:actual_vs_ideal_psnr}
     \end{subfigure}
     \hfill
     \begin{subfigure}[t]{0.49\textwidth}
         \centering
         \includegraphics[width=\textwidth]{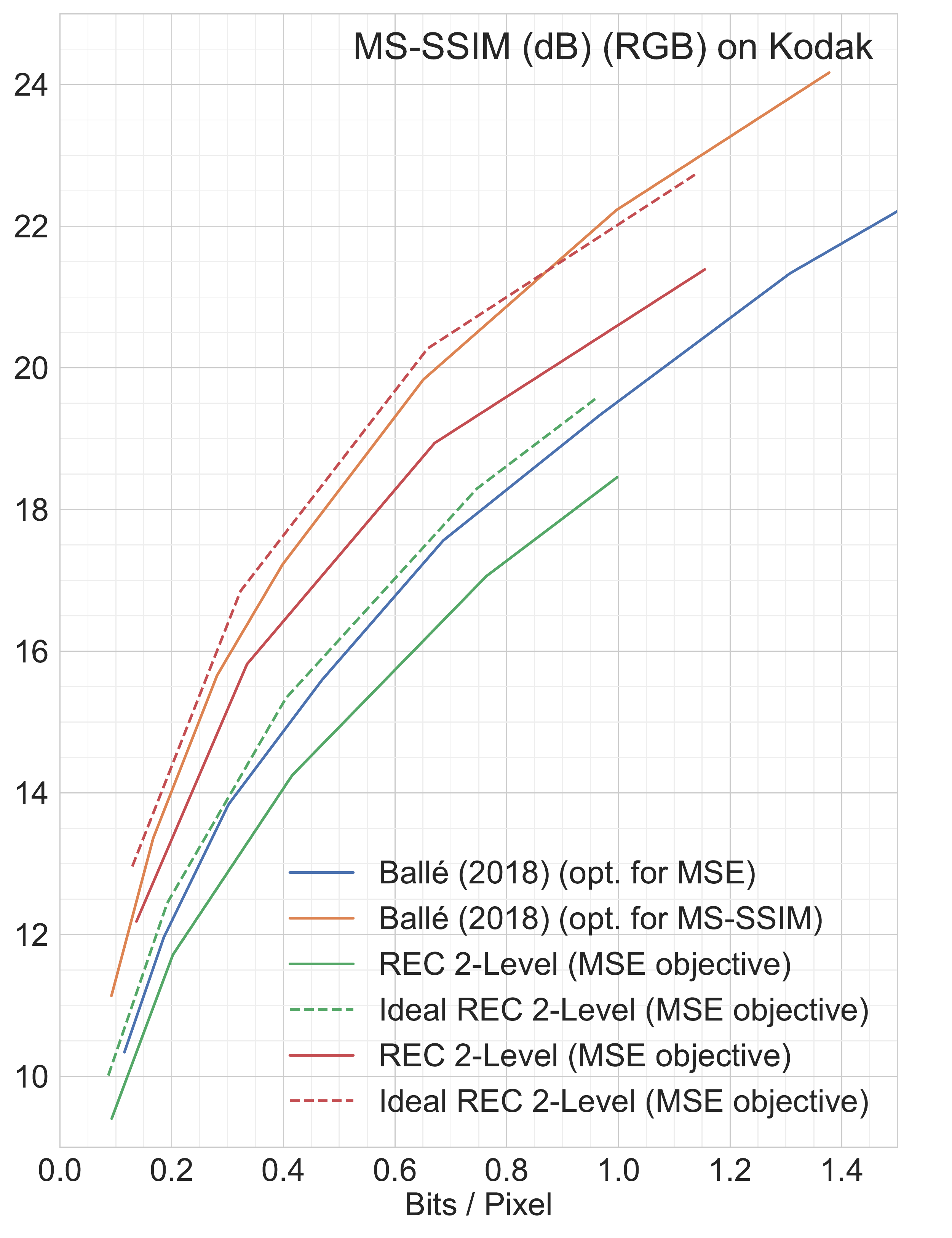}
         \caption{Performance on MS-SSIM.}
         \label{fig:actual_vs_ideal_ms_ssim}
     \end{subfigure}
        \caption{Actual vs Ideal Performance Comparison}
        \label{fig:actual_vs_ideal_comparison}
\end{figure}

We quantify these discrepancies through visualizing \textit{actual} and \textit{ideal} aggregate rate-distortion curves on the Kodak dataset in Figure \ref{fig:actual_vs_ideal_comparison}. Concretely, we calculate the actual performance as in the main text, i.e. the bits per pixel are simply calculated from the compressed file size, and the distortion is calculated using the slightly biased sample given by iREC. The ideal bits per pixel is calculated by dividing the $\KLqp$ by the number of pixels, and the ideal distortion is calculated using a sample drawn from $q(\rvz \mid \vx)$.
\par
As we can see, the distortion gap increases in low-distortion regimes. This is unsurprising, since a low-distortion model's decoder will be more sensitive to biased samples. Interestingly, the ideal performance of our model matches the performance of the method of Ball\'e et al. (2018), even though they used a much more flexible non-parametric hyperprior and their priors and posteriors were picked to suit image compression. On the other hand, our model only used diagonal Gaussian distributions everywhere.
\subsection{Performance Comparisons on Individual Kodak Images}
\par
Aggregate rate-distortion curves can only serve as a way to compare competing methods, and cannot be used to assess absolute method performance. To address this, we present performance comparisons on individual Kodak images juxtaposed with the images and their reconstructions.


\begin{figure}
     \centering
    \begin{subfigure}[t]{0.49\textwidth}
         \includegraphics[width=\textwidth]{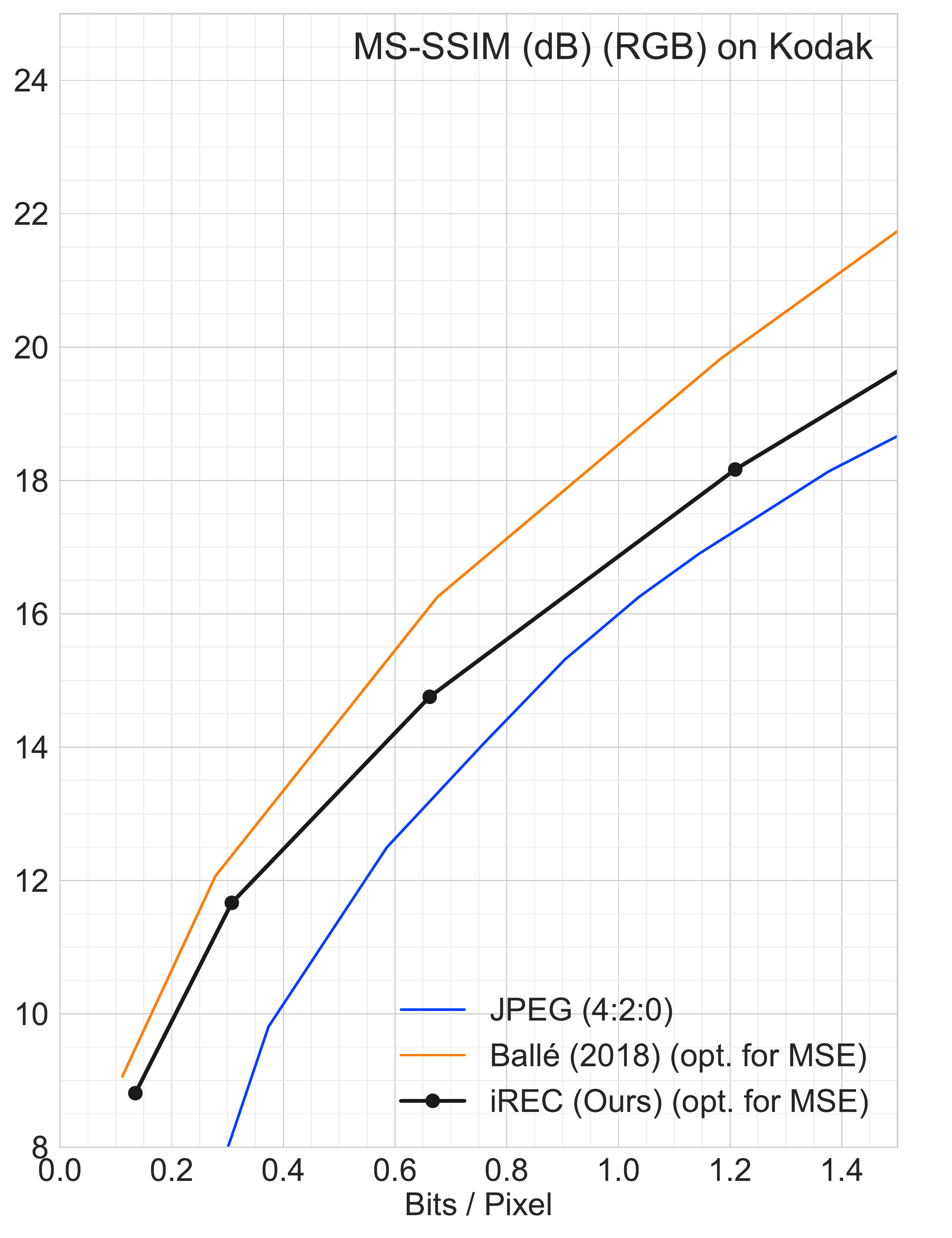} 
         \caption{MS-SSIM rate-distortion curve for Kodak image 1.}
     \end{subfigure}
     \begin{subfigure}[t]{0.49\textwidth}
         \includegraphics[width=\textwidth]{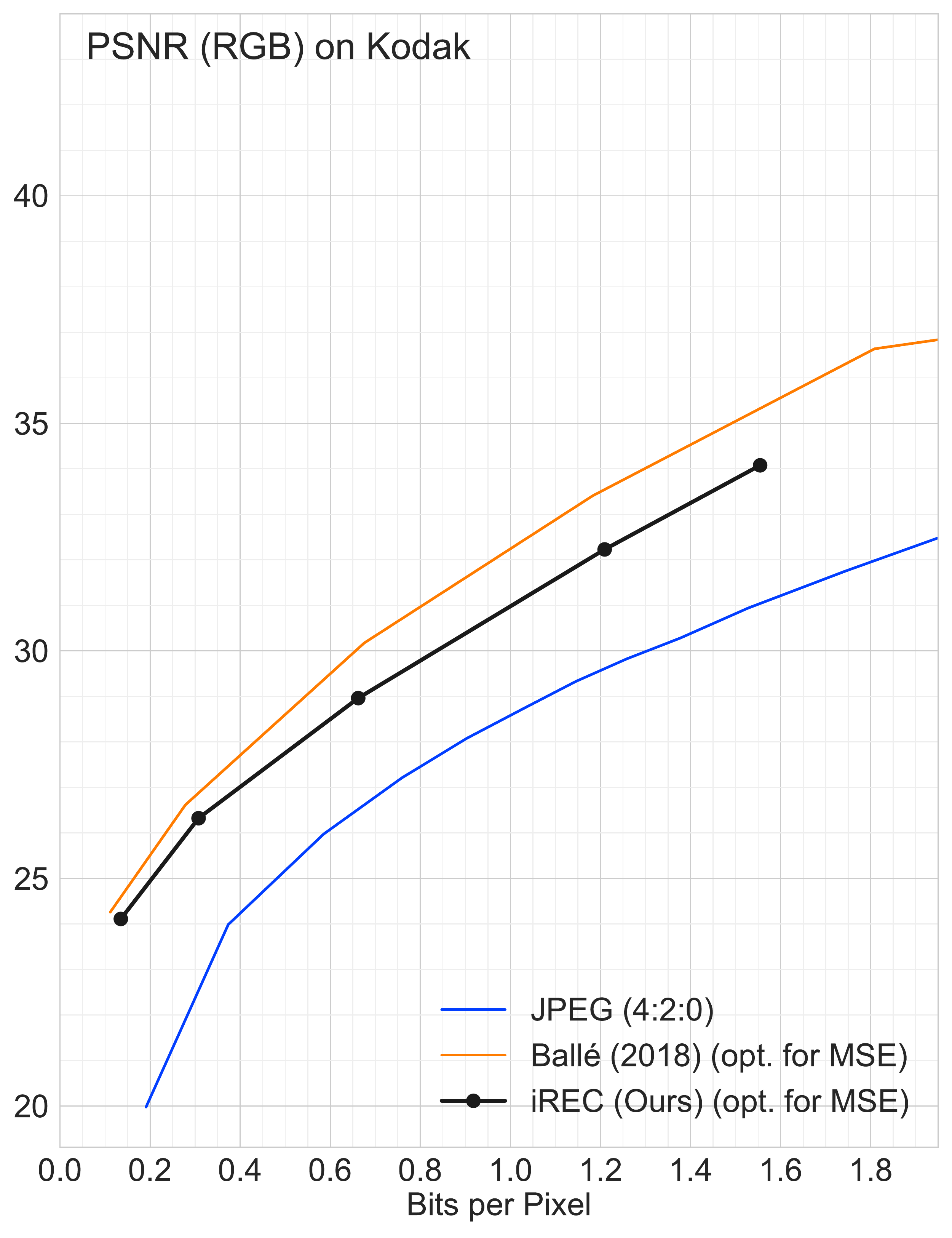} 
         \caption{PSNR rate-distortion curve for Kodak image 1.}
     \end{subfigure}
     \vspace{0.3cm}
     \begin{subfigure}[t]{0.49\textwidth}
         \centering
         \vspace{0.3cm}
         \includegraphics[width=\textwidth]{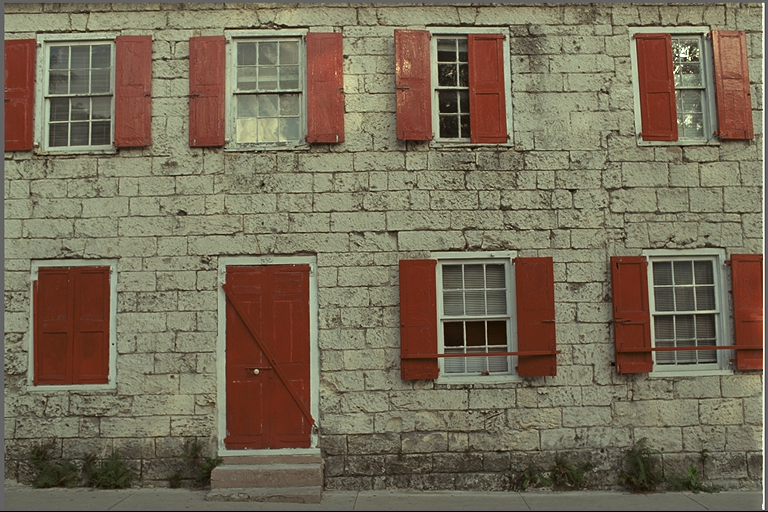}
         \caption{Kodak image 1: Original}
         \label{fig:rec_lossless_compression}
     \end{subfigure}
     \hfill
     \begin{subfigure}[t]{0.49\textwidth}
         \centering
         \vspace{0.3cm}
         \includegraphics[width=\textwidth]{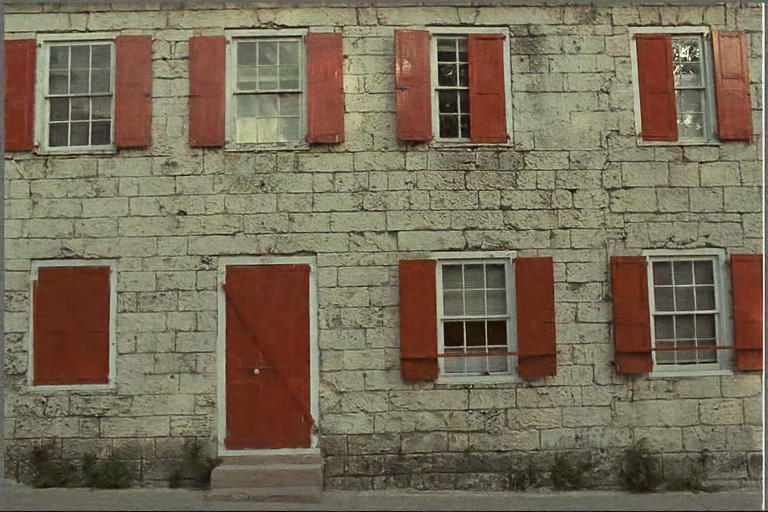}
         \caption{Kodak image 1: Reconstruction using model trained with $\lambda = 0.01$}
     \end{subfigure}
\end{figure}


\begin{figure}
     \centering
    \begin{subfigure}[t]{0.49\textwidth}
         \includegraphics[width=\textwidth]{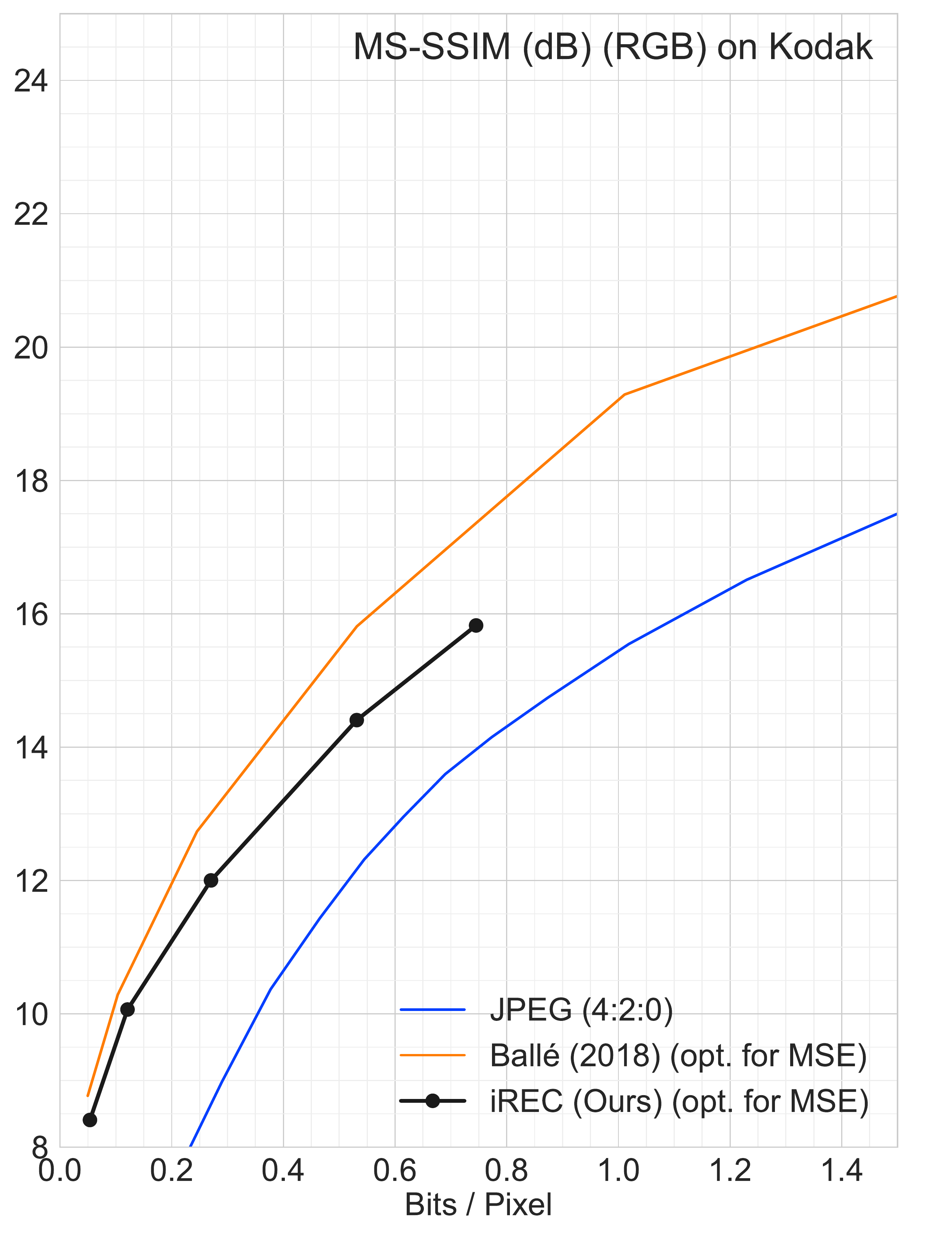} 
         \caption{MS-SSIM rate-distortion curve for Kodak image 2.}
     \end{subfigure}
     \begin{subfigure}[t]{0.49\textwidth}
         \includegraphics[width=\textwidth]{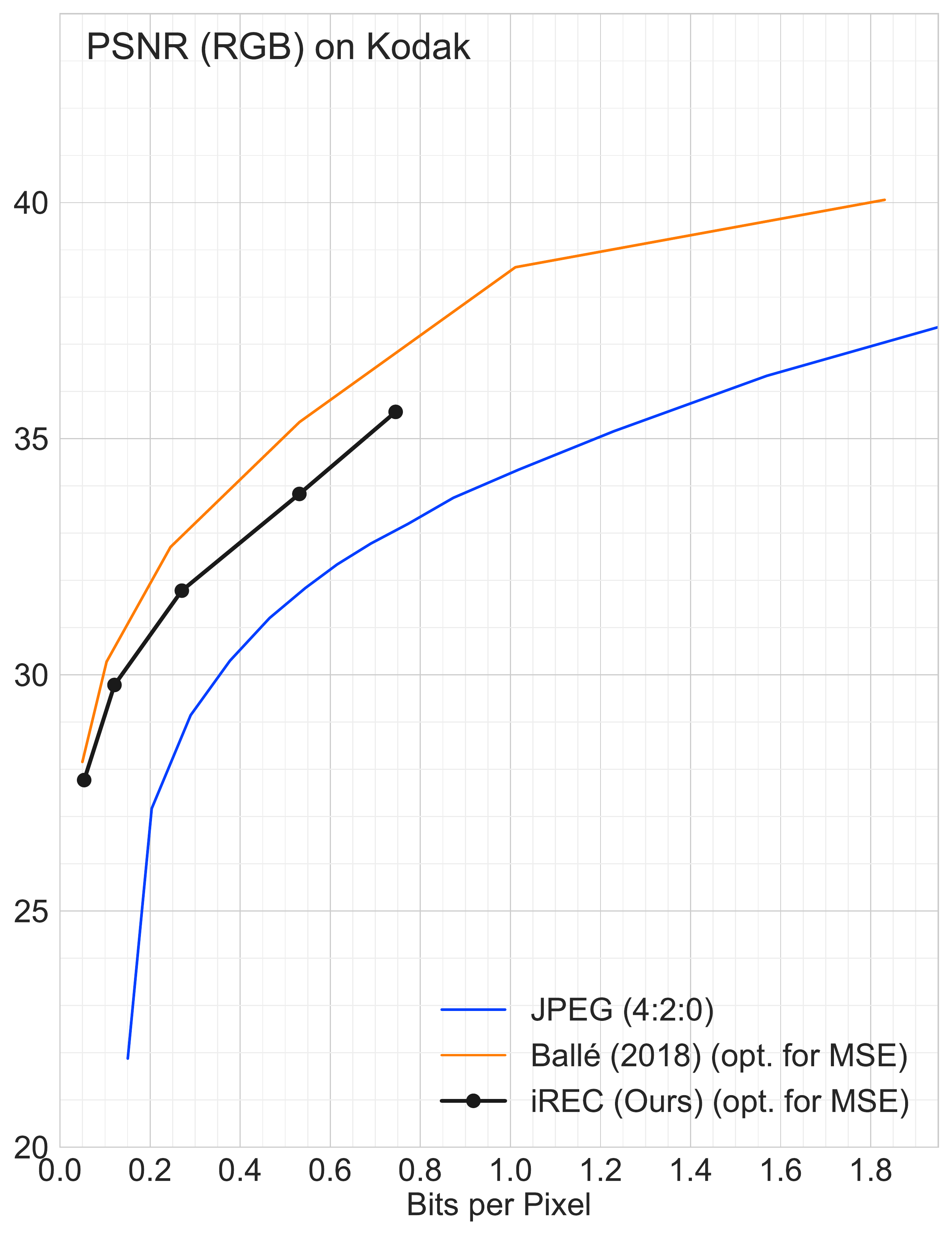} 
         \caption{PSNR rate-distortion curve for Kodak image 2.}
     \end{subfigure}
     \vspace{0.3cm}
     \begin{subfigure}[t]{0.49\textwidth}
         \centering
         \vspace{0.3cm}
         \includegraphics[width=\textwidth]{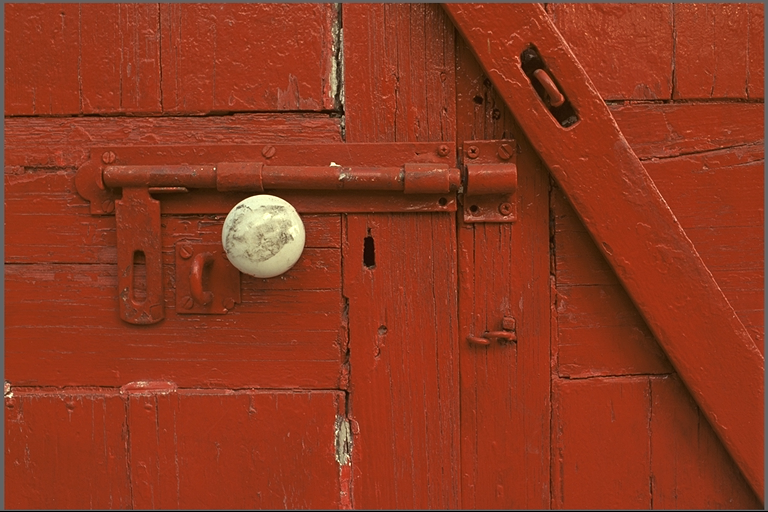}
         \caption{Kodak image 2: Original}
         \label{fig:rec_lossless_compression}
     \end{subfigure}
     \hfill
     \begin{subfigure}[t]{0.49\textwidth}
         \centering
         \vspace{0.3cm}
         \includegraphics[width=\textwidth]{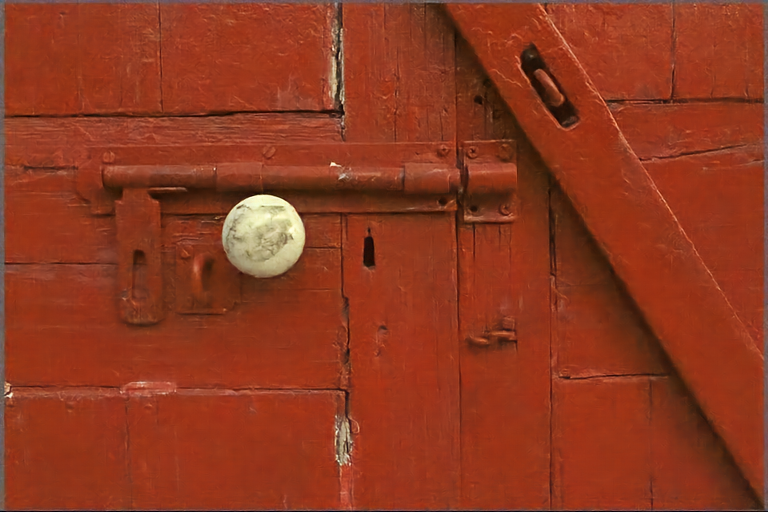}
         \caption{Kodak image 2: Reconstruction using model trained with $\lambda = 0.01$}
     \end{subfigure}
\end{figure}


\begin{figure}
     \centering
    \begin{subfigure}[t]{0.49\textwidth}
         \includegraphics[width=\textwidth]{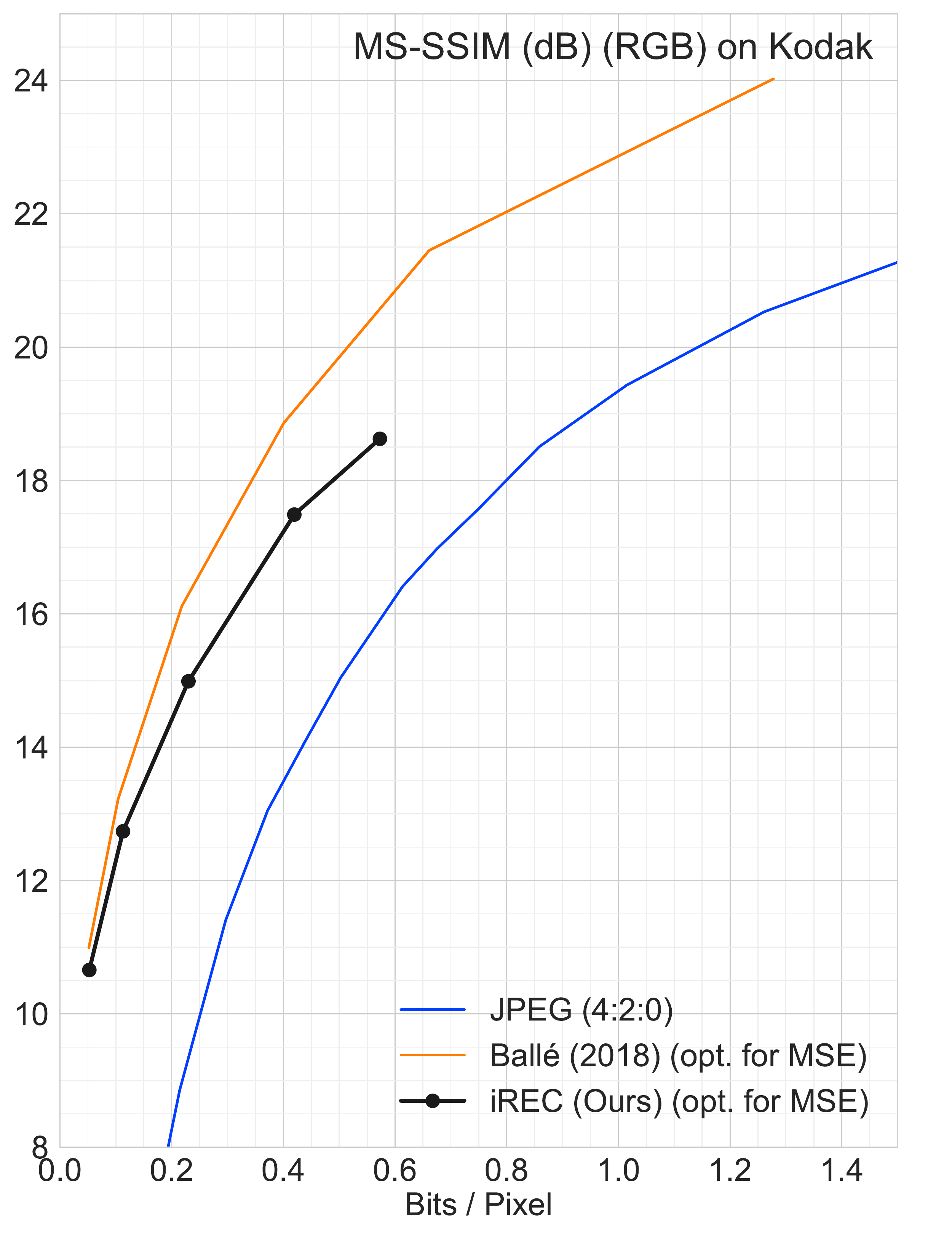} 
         \caption{MS-SSIM rate-distortion curve for Kodak image 3.}
     \end{subfigure}
     \begin{subfigure}[t]{0.49\textwidth}
         \includegraphics[width=\textwidth]{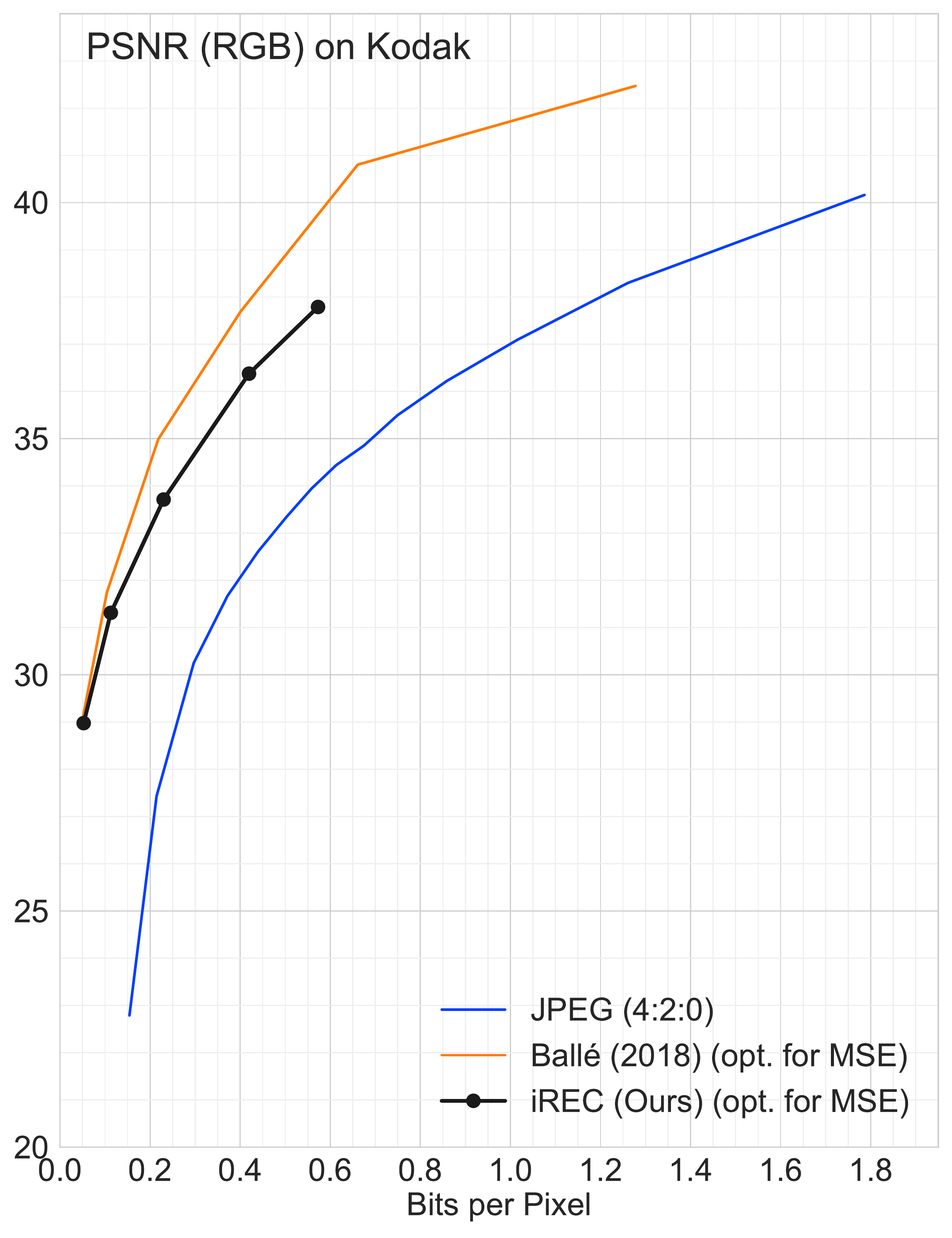} 
         \caption{PSNR rate-distortion curve for Kodak image 3.}
     \end{subfigure}
     \vspace{0.3cm}
     \begin{subfigure}[t]{0.49\textwidth}
         \centering
         \vspace{0.3cm}
         \includegraphics[width=\textwidth]{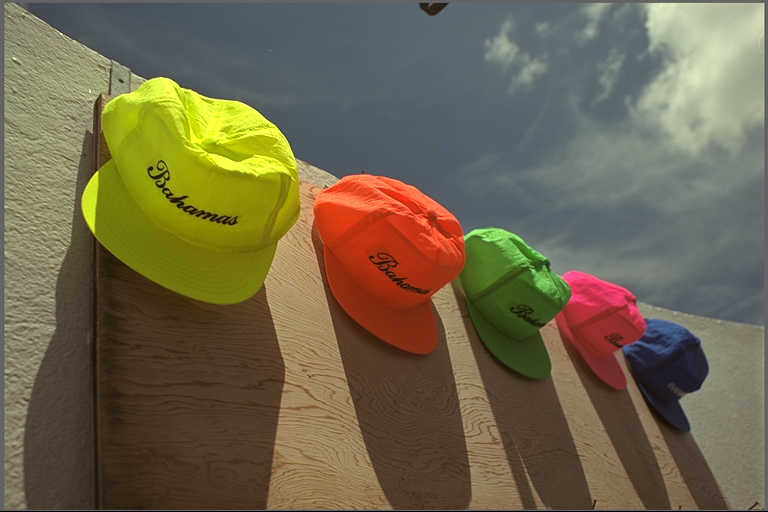}
         \caption{Kodak image 3: Original}
         \label{fig:rec_lossless_compression}
     \end{subfigure}
     \hfill
     \begin{subfigure}[t]{0.49\textwidth}
         \centering
         \vspace{0.3cm}
         \includegraphics[width=\textwidth]{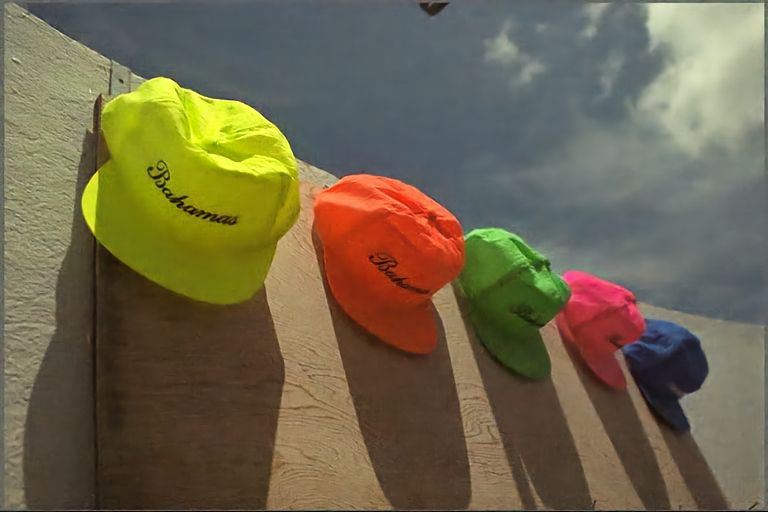}
         \caption{Kodak image 3: Reconstruction using model trained with $\lambda = 0.01$}
     \end{subfigure}
\end{figure}


\begin{figure}
     \centering
    \begin{subfigure}[t]{0.49\textwidth}
         \includegraphics[width=\textwidth]{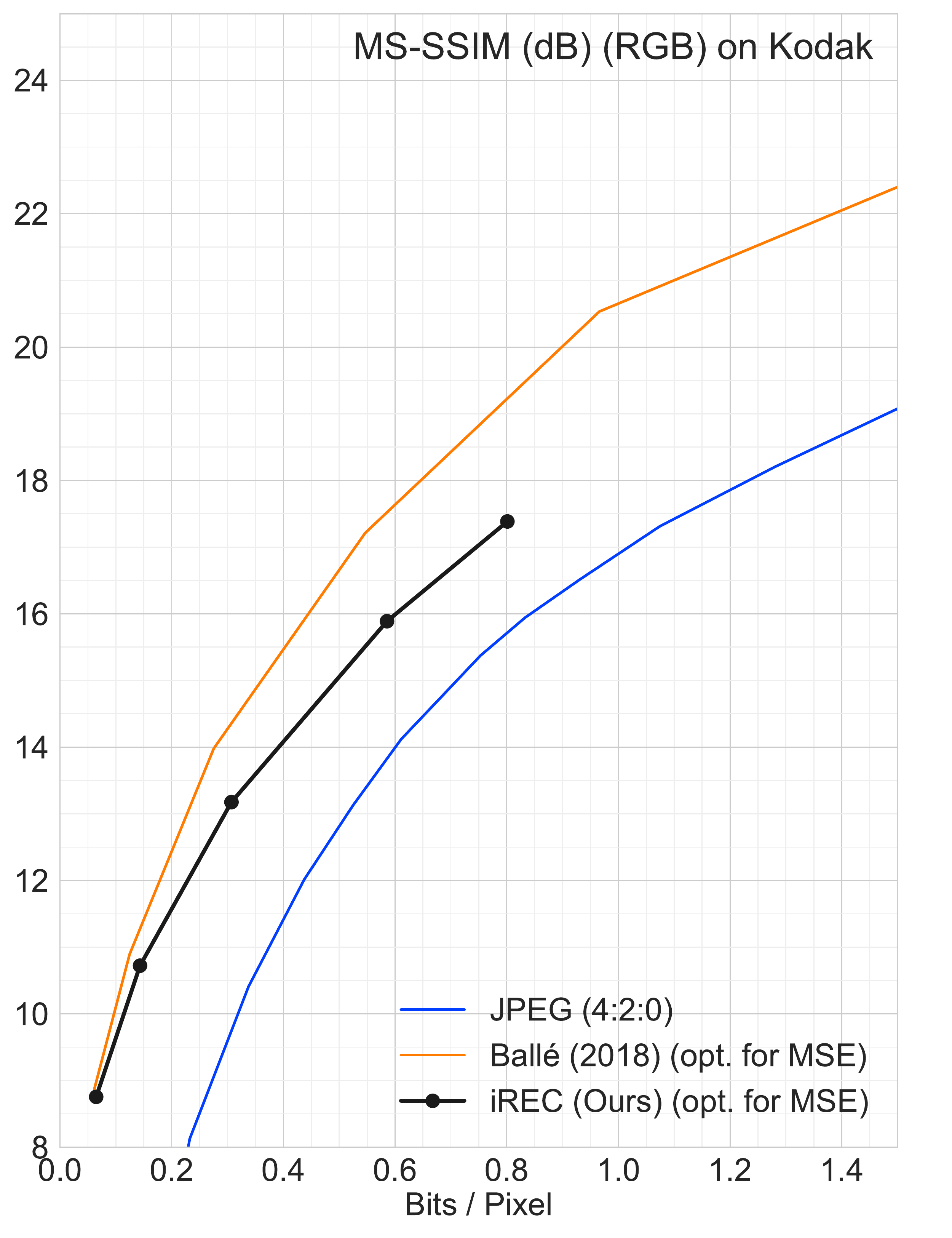} 
         \caption{MS-SSIM rate-distortion curve for Kodak image 4.}
     \end{subfigure}
     \begin{subfigure}[t]{0.49\textwidth}
         \includegraphics[width=\textwidth]{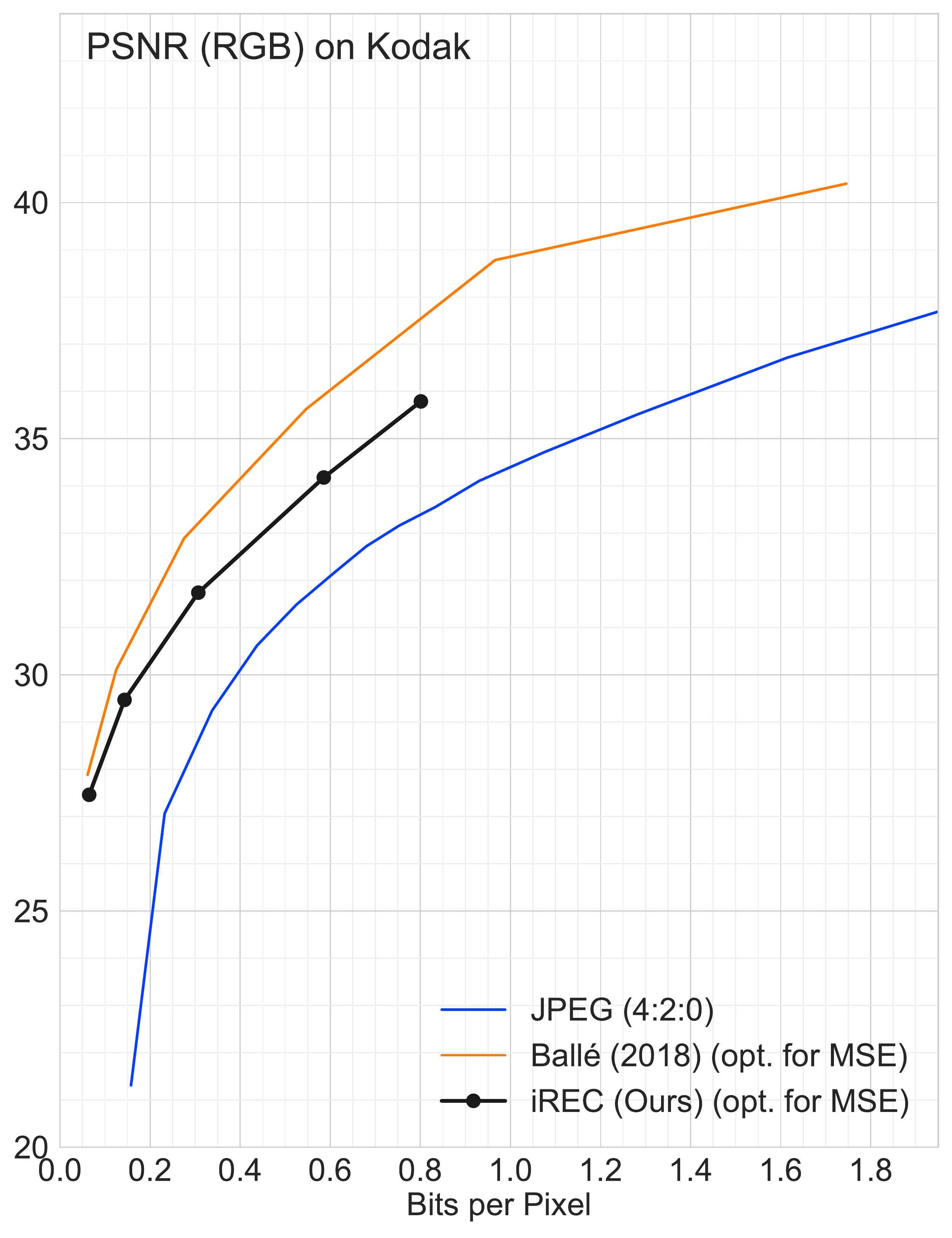} 
         \caption{PSNR rate-distortion curve for Kodak image 4.}
     \end{subfigure}
     \vspace{0.3cm}
     \begin{subfigure}[t]{0.49\textwidth}
         \centering
         \vspace{0.3cm}
         \includegraphics[width=\textwidth]{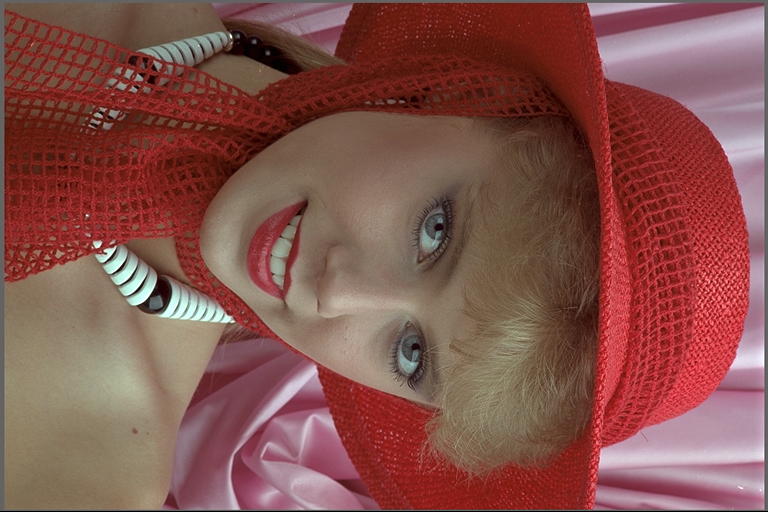}
         \caption{Kodak image 4: Original}
         \label{fig:rec_lossless_compression}
     \end{subfigure}
     \hfill
     \begin{subfigure}[t]{0.49\textwidth}
         \centering
         \vspace{0.3cm}
         \includegraphics[width=\textwidth]{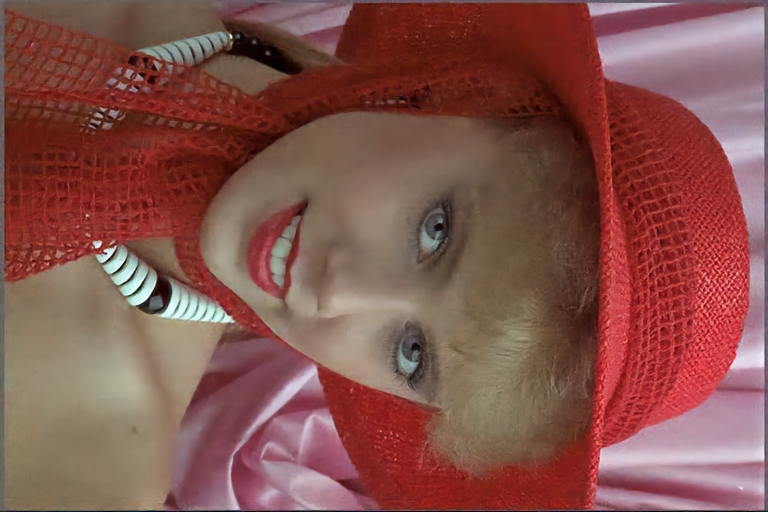}
         \caption{Kodak image 4: Reconstruction using model trained with $\lambda = 0.01$}
     \end{subfigure}
\end{figure}

\end{document}